\pdfoutput=1

\documentclass[aps,prd,twocolumn,preprintnumbers,superscriptaddress,nofootinbib,floatfix]{revtex4-1}

\usepackage{amsmath,amssymb}
\usepackage{bm}
\usepackage{graphicx}
\usepackage{epstopdf}
\usepackage{hyperref}
\usepackage{array}
\usepackage[utf8]{inputenc}
\usepackage{soul}
\usepackage[usenames, dvipsnames]{color}

\usepackage{slashed}
\usepackage{tabularx}
\newcolumntype{L}[1]{>{\raggedright\let\newline\\\arraybackslash\hspace{0pt}}m{#1}}
\newcolumntype{C}[1]{>{\centering\let\newline\\\arraybackslash\hspace{0pt}}m{#1}}
\newcolumntype{R}[1]{>{\raggedleft\let\newline\\\arraybackslash\hspace{0pt}}m{#1}}

\newcommand{\vect}{\boldsymbol}

\newcommand{\ccEps}{Mg(SO$_4)\!\cdot\!7($H$_2$O)}

\newcommand{\ccHal}{NaCl}

\newcommand{\ccNch}{Mn$^{2+}_2$SiO$_3$(OH)$_2 \! \cdot \! ($H$_2$O)}

\newcommand{\ccOli}{Mg$_{1.6}$Fe$^{2+}_{0.4}$(SiO$_4$)}

\usepackage{xcolor}

\begin{document}

\title{Paleo-Detectors for Galactic Supernova Neutrinos} 

\newcommand{\OKC}{\affiliation{The Oskar Klein Centre for Cosmoparticle Physics, Department of Physics,\\Stockholm University, Alba Nova, 10691 Stockholm, Sweden}}
\newcommand{\Nordita}{\affiliation{Nordita, KTH Royal Institute of Technology and Stockholm University, Roslagstullsbacken 23, 10691 Stockholm, Sweden}}
\newcommand{\GRAPPA}{\affiliation{Gravitation Astroparticle Physics Amsterdam (GRAPPA), Institute for Theoretical Physics Amsterdam and \\ Delta Institute for Theoretical Physics, University of Amsterdam, Science Park 904, 1098 XH Amsterdam, The Netherlands}}
\newcommand{\LCTP}{\affiliation{Leinweber Center for Theoretical Physics, University of Michigan, Ann Arbor, MI 48109, USA}}
\newcommand{\NCBJ}{\affiliation{National Centre for Nuclear Research, 05-400 Otwock, \'{S}wierk, Poland}}
\newcommand{\UTexas}{\affiliation{Department of Physics, University of Texas, Austin, TX 78712, USA}}
\newcommand{\SITP}{\affiliation{Stanford Institute for Theoretical Physics, Department of Physics, Stanford University, Stanford, CA 94305, USA}}

\author{Sebastian~Baum}
\email{sbaum@stanford.edu}
\OKC
\Nordita
\SITP

\author{Thomas~D.~P.~Edwards}
\email{t.d.p.edwards@uva.nl}
\GRAPPA

\author{Bradley~J.~Kavanagh}
\email{b.j.kavanagh@uva.nl}
\GRAPPA

\author{Patrick~Stengel}
\email{patrick.stengel@fysik.su.se}
\OKC

\author{Andrzej~K.~Drukier}
\email{adrukier@gmail.com}
\OKC

\author{Katherine~Freese}
\email{ktfreese@umich.edu}
\OKC
\Nordita
\LCTP
\UTexas

\author{Maciej~G\'{o}rski}
\email{maciej.gorski@ncbj.gov.pl}
\NCBJ

\author{Christoph~Weniger}
\email{c.weniger@uva.nl}
\GRAPPA

\begin{abstract}
Paleo-detectors are a proposed experimental technique in which one would search for traces of recoiling nuclei in ancient minerals. Natural minerals on Earth are as old as $\mathcal{O}(1)\,$Gyr and, in many minerals, the damage tracks left by recoiling nuclei are also preserved for timescales long compared to $1\,$Gyr once created. Thus, even reading out relatively small target samples of order $100\,$g, paleo-detectors would allow one to search for very rare events thanks to the large exposure, $\varepsilon \sim 100\,{\rm g}\,{\rm Gyr} = 10^5\,{\rm t}\,{\rm yr}$. Here, we explore the potential of paleo-detectors to measure nuclear recoils induced by neutrinos from galactic core collapse supernovae. We find that they would not only allow for a direct measurement of the average core collapse supernova rate in the Milky Way, but would also contain information about the time-dependence of the local supernova rate over the past $\sim 1\,$Gyr. Since the supernova rate is thought to be directly proportional to the star formation rate, such a measurement would provide a determination of the local star formation history. We investigate the sensitivity of paleo-detectors to both a smooth time evolution and an enhancement of the core collapse supernova rate on relatively short timescales, as would be expected for a starburst period in the local group.
\end{abstract}

\maketitle

\section{Introduction}

Supernovae (SNe) play an important role in cosmology and astrophysics. For example, SN feedback is thought to be an important ingredient for understanding galaxy formation~\cite{Dekek:1986gu}. While many extragalactic SNe have been observed~\cite{Akerlof:2002xu,Law:2009ys,Leaman:2010kb,Graur:2011,Graur:2013msa,Sako:2014qmj,Krisciunas:2017yoe}, allowing for a rather precise determination of the cosmic SN rate~\cite{Madau:2014bja,Strolger:2015kra,Petrushevska:2016kie}, only a handful of SNe have been observed in the local group~\cite{Arnett:1990au,Green:2003ir}. To date, no direct measurement of the SN rate in the Milky Way exists; estimates in the literature suggest a rate of a few SNe per century~\cite{Cappellaro:2003eg,Diehl:2006cf,Strumia:2006db,Li:2011,Botticella:2011nd,Adams:2013ana}. 

In this paper, we explore the potential of {\it paleo-detectors} to measure the core collapse (CC) SN rate in our galaxy. Paleo-detectors have recently been studied as a method for the direct detection of dark matter~\cite{Baum:2018tfw,Drukier:2018pdy,Edwards:2018hcf}. In certain minerals, e.g.~those long used as solid state track detectors, recoiling nuclei leave damage tracks~\cite{Fleischer:1964,Fleischer383,Fleischer:1965yv,GUO2012233}. Once created, such tracks are preserved over geological timescales. In paleo-detectors, one would search for damage tracks in minerals as old as $\sim 1\,$Gyr using modern nano-technology such as helium-ion beam or X-ray microscopy~\cite{Baum:2018tfw,Drukier:2018pdy}; see also Refs.~\cite{Goto:1958,Goto:1963zz,Fleischer:1969mj,Fleischer:1970zy,Fleischer:1970vm,Alvarez:1970zu,Kolm:1971xb,Eberhard:1971re,Ross:1973it,Price:1983ax,Kovalik:1986zz,Ghosh:1990ki,Jeon:1995rf,Collar:1999md,SnowdenIfft:1995ke,Collar:1994mj,Engel:1995gw,SnowdenIfft:1997hd} for related earlier ideas that use ancient minerals to probe rare events.

Besides probing dark matter, paleo-detectors would also detect neutrinos via nuclear recoils induced by coherent neutrino-nucleus scattering. Thus, paleo-detectors could, for the first time, provide a direct measurement of the galactic CC SN rate over the past $\sim 1\,$Gyr, as initially proposed in Ref.~\cite{Collar:1995aw}.

SNe are broadly divided into thermonuclear (type Ia) and CC SNe. Only the latter ones are expected to produce a significant flux in neutrinos. The progenitors of CC SNe are massive stars (heavier than $\sim 8\,M_\odot$). Such stars are short-lived, with lifetimes $\lesssim 50\,$Myr, see e.g. Refs.~\cite{Hirschi:2004ks,Smartt:2009zr}. Thus, on the timescales relevant for paleo-detectors (order $100\,$Myr and longer), the CC SN rate closely traces the star formation rate, see e.g. Refs.~\cite{Madau:1998dg,Dahlen:2004km,Madau:2014bja,Strolger:2015kra}. Considerable uncertainties exist in the estimates of the local star formation rate, see e.g. Refs.~\cite{Cignoni:2006hm,Czekaj:2014kla,Snaith:2015,Haywood:2016,Mor:2019} for recent work. A direct measurement of the galactic CC SN rate would thus provide valuable information for understanding our galaxy.

While paleo-detectors would only provide a coarse-grained time resolution, we demonstrate that some time-dependent information of the galactic CC SN rate can still be obtained. We consider two distinct cases: (i) we study how well a smooth time evolution of the CC SN rate could be constrained by paleo detectors, and (ii) we investigate if paleo-detectors could be used to find evidence for a starburst period in the Milky Way within the last $\sim 1\,$Gyr. Both of these cases would provide information about the star formation history of the Milky Way. SN explosions in close proximity to Earth have also been hypothesized to give rise to mass extinction events~\cite{Ellis:1993kc,Ellis:1995qb,Fields:1998hd,Svensmark:2012,Thomas:2016fcp,Melott:2017blc,Melott:2017met,Melott:2019fxa,Fields:2019jtj}. We demonstrate that paleo-detectors could probe a single close-by CC SN explosion if it occurred during the exposure time.

The remainder of this paper is organized as follows. In Sec.~\ref{sec:Sig}, we discuss the track length spectrum produced in paleo-detectors from galactic CC SN neutrinos. In Sec.~\ref{sec:Bkg}, we briefly review the relevant sources of backgrounds; a more detailed discussion can be found in Ref.~\cite{Drukier:2018pdy}. The best read-out technique for the damage tracks induced by galactic CC SN neutrinos appears to be small angle X-ray scattering tomography, which we discuss in Sec.~\ref{sec:ReadOut}. Our projections for the sensitivity of paleo-detectors to galactic CC SNe as well as the time-evolution of the CC SN rate are discussed in Sec.~\ref{sec:Res}. In Sec.~\ref{sec:Discussion} we summarize and discuss. Appendices~\ref{app:Uranium} and~\ref{app:Stat} contain additional details about uranium-238 concentrations in typical target materials and the statistical techniques used in this work, respectively. All relevant code can be found online at \href{https://doi.org/10.5281/zenodo.3066206}{DOI:10.5281/zenodo.3066206}.

\section{Galactic Core Collapse Supernova Signal} \label{sec:Sig}

We begin by describing the predicted flux of neutrinos from CC SNe and the corresponding signal which we expect to be produced in paleo-detectors.

CC SNe are amongst the brightest astrophysical sources of neutrinos. In fact, SN~1987A (which occurred in the Large Magellanic Cloud) is the only astrophysical object, besides the Sun (and the recently-claimed flaring blazar TXS~0506+056~\cite{IceCube:2018dnn}), to be directly observed in neutrinos. Despite the important role neutrinos play in SN explosions~\cite{Langer:2012jz,Foglizzo:2015dma,Muller:2017wgc}, the precise shape and normalization of the emitted neutrino spectra are poorly understood. The only experimental knowledge stems from the emission of SN~1987A: the 20 events observed by Kamiokande-II~\cite{Hirata:1987hu}, eight events by IMB~\cite{Bionta:1987qt}, five events by LSD~\cite{Aglietta:1987}, and five events by the Baksan Neutrino Observatory~\cite{Alekseev:1988gp}. Alternatively, neutrino spectra can be predicted from simulations, which are usually well-fitted by a pinched Fermi-Dirac distribution~\cite{Keil:2002in}
\begin{equation}\label{eq:nu_spec}
   \left(\frac{\mathrm{d}n}{\mathrm{d}E}\right)_{\nu_i} = E_\nu^{\rm tot} \frac{\left(1+\alpha\right)^{1+\alpha}}{\Gamma(1+\alpha)} \frac{E^{\alpha}}{\langle E_\nu\rangle^{2+\alpha}} e^{\left[ -\left(1+\alpha\right) \frac{E}{\langle E_\nu \rangle} \right]} \;,
\end{equation}
where $E_\nu^{\rm tot}$ is the energy radiated in the neutrino species $\nu_i$, $\langle E_\nu\rangle$ is the average neutrino energy (approximately given by the core-temperature of the SN), and $\alpha$ is the spectral shape parameter. However, sizable differences remain between parameter values inferred from different simulations, see for example Refs.~\cite{Keil:2002in,Tamborra:2012ac,Mirizzi:2015eza,Horiuchi:2017qja}. Here, we use the values suggested by Ref.~\cite{Horiuchi:2017qja}, listed in Tab.~\ref{tab:nu_params}. 

\begin{table}
   \begin{center}
   \begin{tabular}{L{1cm} C{1.5cm} C{1.5cm} C{1.5cm}}
   \hline\hline
   $\nu$ & $E_\nu^{\rm tot}$ [erg] & $\langle E_\nu \rangle$ [MeV] & $\alpha$ \\
   \hline
   $\nu_e$ & $6 \times 10^{52}$ & $13.3$ & $3.0$ \\
   $\nu_{\bar{e}}$ & $4.3 \times 10^{52}$ & $14.6$ & $3.3$ \\
   $\nu_x$ & $2 \times 10^{52}$ & $15$ & $3$ \\
   \hline\hline
   \end{tabular}
   \caption{Parameters of the neutrino spectra, Eq.~\eqref{eq:nu_spec}, for electron neutrinos, anti-electron neutrinos, and $\nu_x \equiv \{\nu_\mu, \nu_{\bar{\mu}}, \nu_\tau, \nu_{\bar{\tau}} \}$ used in our numerical calculations~\cite{Horiuchi:2017qja}.}
   \label{tab:nu_params}
   \end{center}
\end{table}

The dominant source of neutrino-induced nuclear recoils are (flavor-blind) neutral current interactions. Thus, the relevant neutrino flux is the sum over all neutrino flavors,
\begin{equation} \label{eq:nu_spec_sum}
   \frac{\mathrm{d}n}{\mathrm{d}E_\nu} = \left(\frac{\mathrm{d}n}{\mathrm{d}E}\right)_{\nu_e} + \left(\frac{\mathrm{d}n}{\mathrm{d}E}\right)_{\nu_{\bar{e}}} + 4 \left(\frac{\mathrm{d}n}{\mathrm{d}E}\right)_{\nu_x} \;,
\end{equation}
where $\nu_x \equiv \{\nu_\mu, \nu_{\bar{\mu}}, \nu_\tau, \nu_{\bar{\tau}} \}$. Since neutral current interactions are flavor blind, we do not need to account for flavor oscillations. These are a major source of uncertainty when calculating the neutrino fluxes from SNe, due to the sizable matter effects in the SN environment. 

The time-averaged neutrino spectrum from Galactic CC SNe at Earth is obtained by integrating over the probability density $f(R_E)$ describing the likelihood for a CC SN to occur at a distance $R_E$ from Earth,
\begin{equation} \label{eq:CCdist_galactocentric}
   \left( \frac{\mathrm{d}\phi}{\mathrm{d}E_\nu} \right)^{\rm gal} = \dot{N}_{\rm CC}^{\rm gal} \frac{\mathrm{d}n}{\mathrm{d}E_\nu} \int_0^\infty \mathrm{d}R_E \frac{f(R_E)}{4\pi R_E{}^2} \;,
\end{equation}
where $\dot{N}_{\rm CC}^{\rm gal}$ is the galactic CC SN rate.\footnote{In principle, the integral over $R_E$ should be truncated at some distance corresponding to the size of our galaxy. Here, we instead use a probability density $f(R_E)$ which takes into account only CC SNe within the galactic disk of the Milky Way.} To obtain $f(R_E)$, we follow Ref.~\cite{Adams:2013ana} and assume that CC SNe occur predominantly in the stellar disk. In galactocentric cylindrical coordinates, the spatial distribution of CC SNe, $\rho$, can then be modeled by a double exponential
\begin{equation} \label{eq:spatial_dist}
   \rho \propto e^{-R/R_d} e^{-|z|/H} \;,
\end{equation}
where $R$ is the galactocentric radius, $z$ is the height above the galactic mid-plane, and we set the disk parameters to $R_d = 2.9\,$kpc and $H = 95\,$pc~\cite{Adams:2013ana}. From Eq.~\eqref{eq:CCdist_galactocentric}, we obtain the probability density as a function of the distance from Earth $f(R_{E})$ by performing a coordinate transformation to the position of the Sun with galactocentric radius $R_\odot = 8.7$\,kpc and height above the disk $H_\odot = 24\,$pc. Note that the position of the Sun with respect to the galactic center will change over the timescales that paleo-detectors were exposed to neutrinos from galactic CC SNe, $\mathcal{O}(1)\,$Gyr.\footnote{The orbital period of the Sun around the galactic center is $T_\odot \sim 250\,$Myr.} The solar system is thought to follow an approximately circular orbit around the galactic center, oscillating about the galactic disk by $\Delta z_\odot \sim 100\,$pc and oscillating in the galactic plane by $\Delta R_\odot \sim 300\,$pc~\cite{Schoenrich:2010,Kawata:2019}. Modifying the distance of the solar system to the galactic center by such an amount would change the neutrino flux from CC SNe at Earth by $\Delta \phi \lesssim 10\,\%$, an error much smaller than the uncertainty on the galactic CC SN rate. In the following, we therefore neglect corrections to the predicted neutrino flux from the changing position of the solar system.

\begin{figure}
   \includegraphics[width=\linewidth]{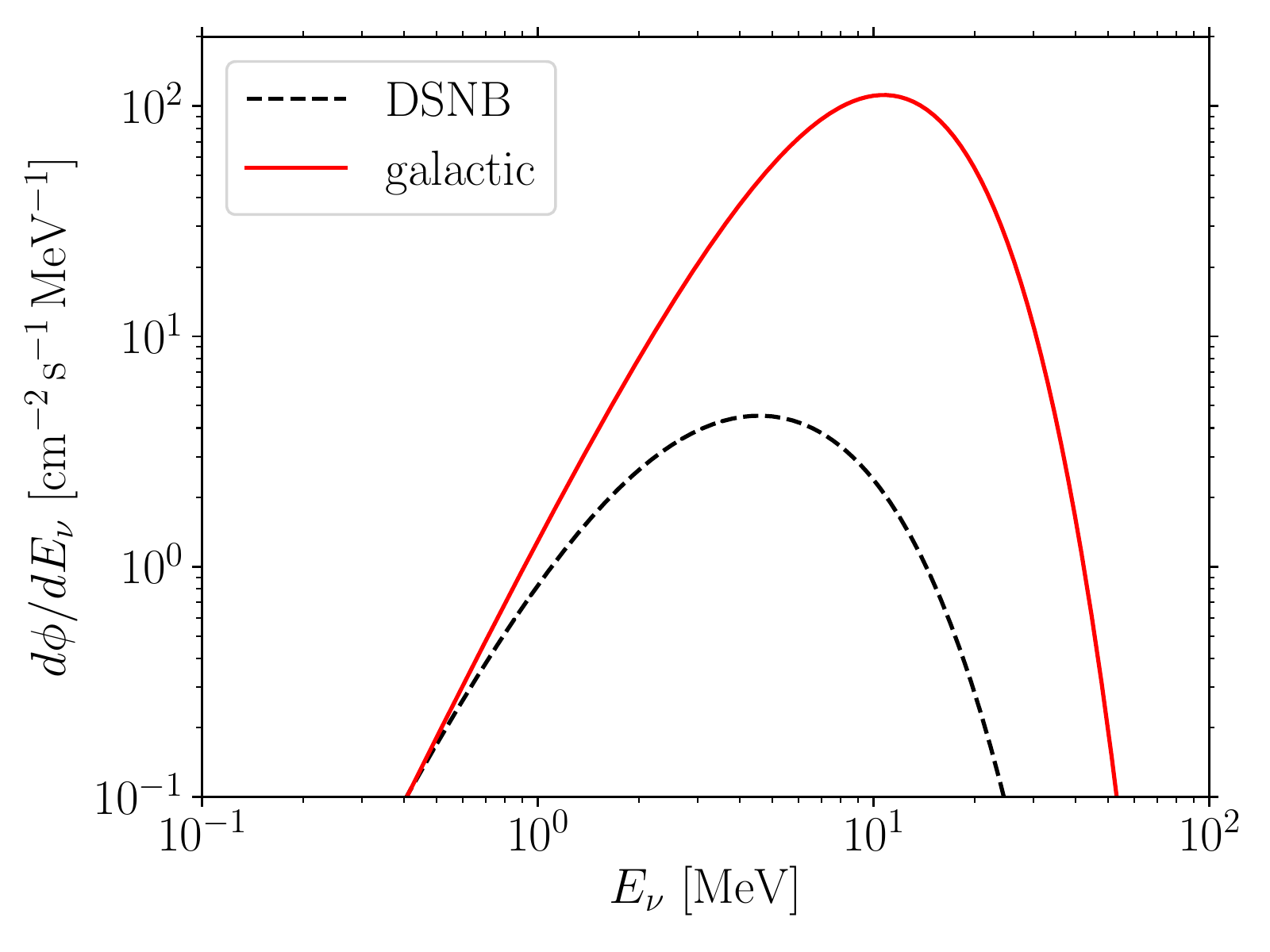}
   \caption{Neutrino flux $\mathrm{d}\phi/\mathrm{d}E_\nu$ (solid red) from galactic CC SNe at Earth as a function of neutrino energy $E_\nu$. Here, we assumed a galactic CC SN rate of $\dot{N}_{\rm CC}^{\rm gal} = 2.3 \times 10^{-2}\,{\rm yr}^{-1}$~\cite{Li:2011}, a spatial distribution of CC SN as given in Eq.~\eqref{eq:spatial_dist}, the neutrino spectrum per CC SN from Eqs.~\eqref{eq:nu_spec} and~\eqref{eq:nu_spec_sum} with the parameters from Tab.~\ref{tab:nu_params}, and averaged the neutrino flux over time-scales much longer than the inverse galactic CC SN rate $(\dot{N}_{\rm CC}^{\rm gal})^{-1} \sim 40\,$yr. For comparison, the black dashed line shows the neutrino flux from distant CC SNe throughout the Universe, the so-called Diffuse SN Background (DSNB). See Ref.~\cite{Beacom:2010kk} for the calculation of the DSNB spectrum; we use the parameterization of the cosmic CC SN rate from Ref.~\cite{Strolger:2015kra}.}
   \label{fig:SN_spec}
\end{figure}

In Fig.~\ref{fig:SN_spec} we show the neutrino spectrum from galactic CC SNe together with the neutrino spectrum expected from distant CC SNe throughout the Universe, the so-called Diffuse SN Background (DSNB). We follow Ref.~\cite{Beacom:2010kk} for the calculation of the DSNB flux, using the parameterization from Ref.~\cite{Strolger:2015kra} for the cosmic CC SN rate, see also Ref.~\cite{Madau:2014bja}. Assuming a galactic CC SN rate of $\dot{N}_{\rm CC}^{\rm gal} = 2.3 \times 10^{-2}\,{\rm yr}^{-1}$~\cite{Li:2011}, we find that the time-averaged neutrino flux from galactic CC SN at Earth peaks at $\mathrm{d}\phi/\mathrm{d}E_\nu \sim 10^2\,{\rm cm}^{-2}\,{\rm s}^{-1}\,{\rm MeV}^{-1}$ with $E_\nu \sim 10\,$MeV. Note that the flux is approximately 100 times that of the DSNB flux. Further, the DSNB spectrum is shifted to lower energies by approximately a factor of two. This shift is due to the peak cosmic CC SN rate occurring at a redshift of $z \sim 1$~\cite{Strolger:2015kra}. Note that estimates of the CC SN rate inferred from the cosmic star formation rate peak at somewhat larger redshifts of $z \sim 2$~\cite{Madau:2014bja}. The DSNB neutrino spectrum obtained from such parameterizations of the CC SN rate would be shifted to even smaller energies than that shown in Fig.~\ref{fig:SN_spec}. However, as we will see in Sec.~\ref{sec:Bkg}, such uncertainties on the DSNB neutrino flux are not important for this work as the dominant background for the signal from galactic CC SNe stems from radiogenic neutrons.

\begin{figure*}
   \includegraphics[width=0.49\linewidth]{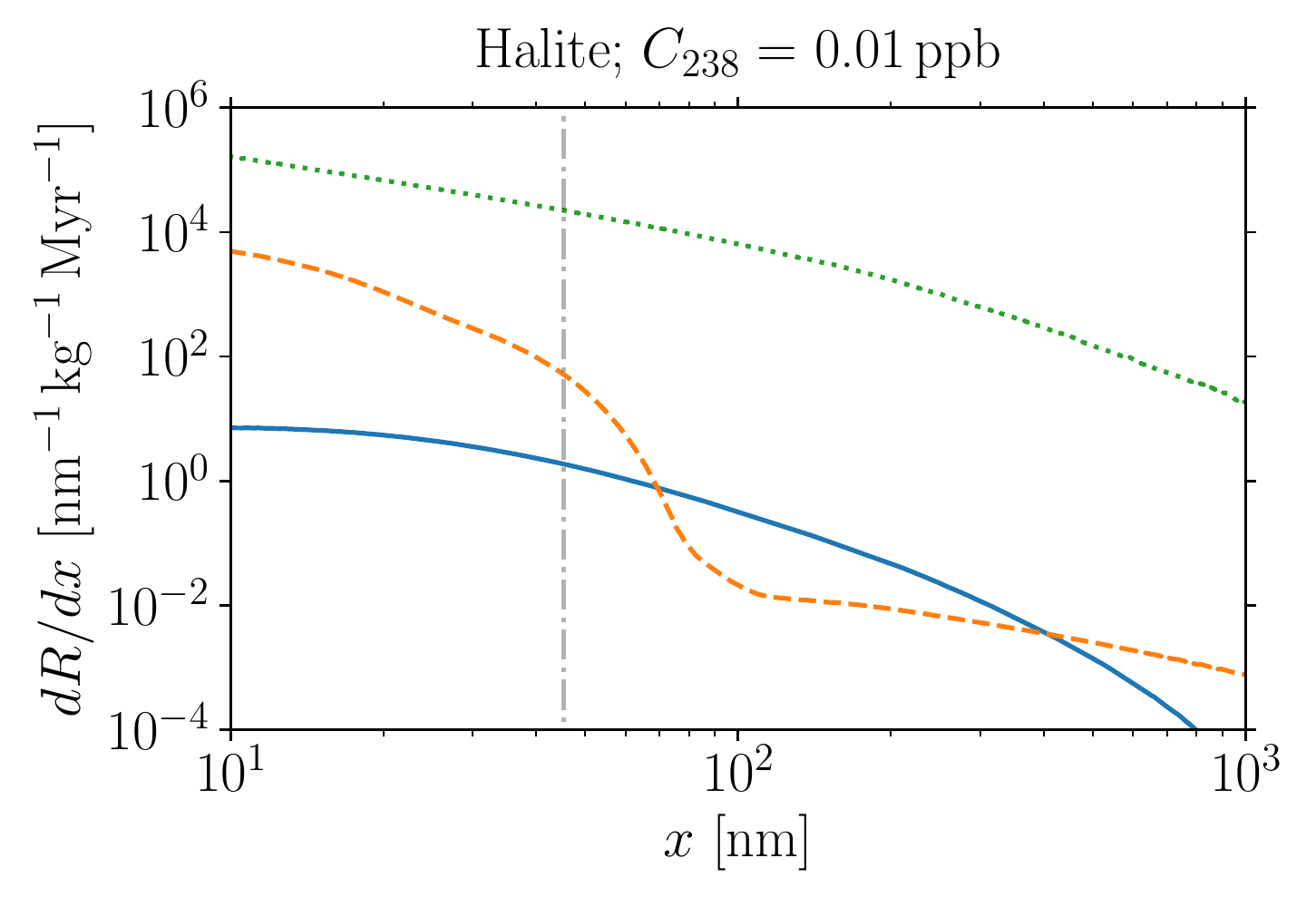}
   \includegraphics[width=0.49\linewidth]{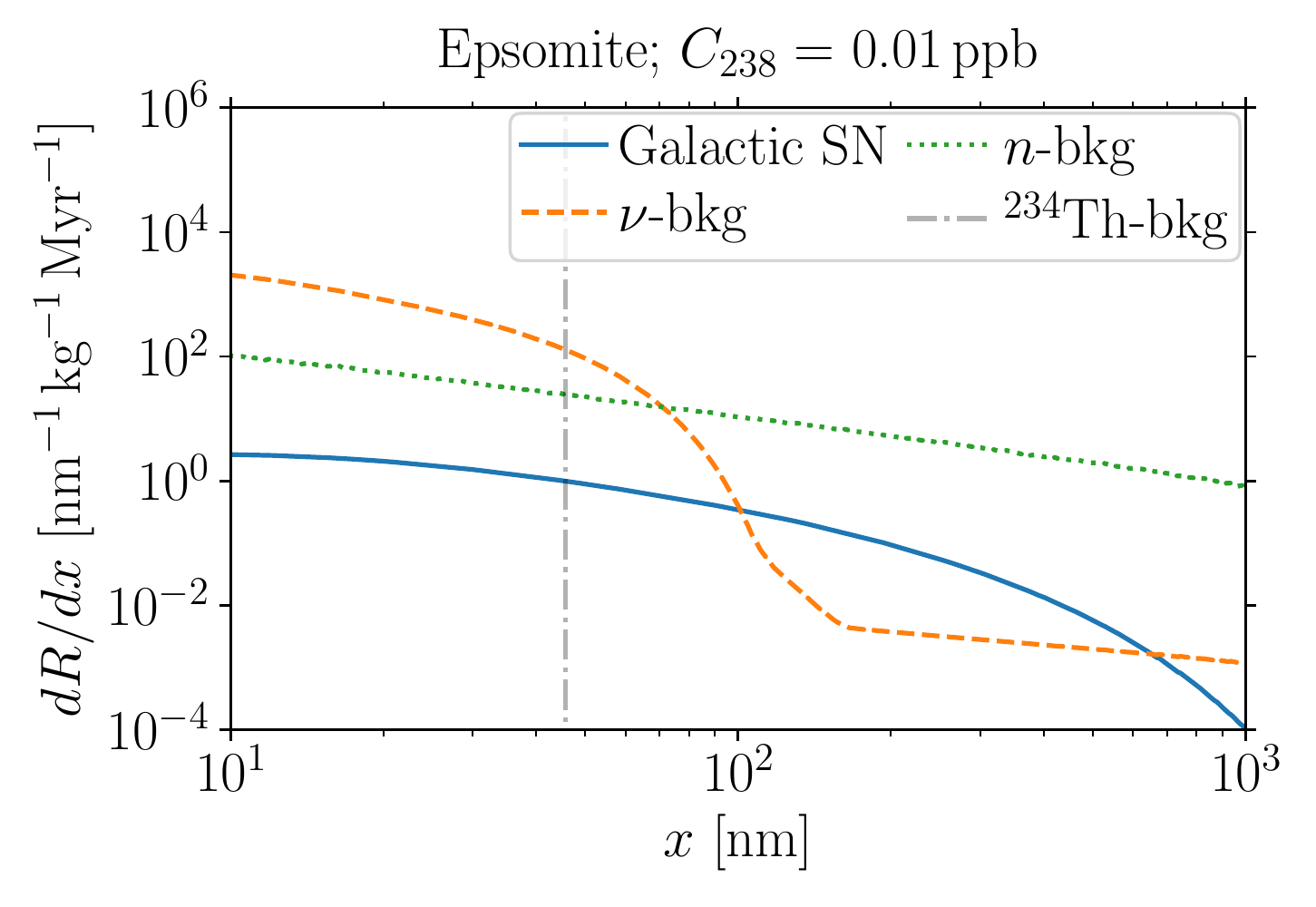}
   \caption{Track length spectra in halite (\ccHal; left) and epsomite [\ccEps; right]. In each panel, the blue solid line shows the spectrum from galactic CC SNe, assuming a rate of $\dot{N}_{\rm CC}^{\rm gal} = 2.3 \times 10^{-2}\,$yr$^{-1}$. The dashed orange line indicates the background spectrum induced by coherent scattering of neutrinos from the Sun, the atmosphere, and the DSNB flux, and the dotted green line shows the background spectrum induced by neutrons from spontaneous fission and $(\alpha,n)$ processes in the target material. The vertical gray dash-dotted lines indicate the track length of the $E_R = 72\,$keV $^{234}$Th nuclei from ($^{238}{\rm U} \to \alpha + {^{234}{\rm Th}}$) decays. See Sec.~\ref{sec:Bkg} for a discussion of the background spectra. For both target minerals, we assumed uranium-238 concentrations of $C_{238} = 0.01\,$ppb by weight. Note that although the signal rate is smaller than the background rate for all track lengths, this does not imply that the signal cannot be measured, see Sec.~\ref{sec:Res}.}
   \label{fig:dRdx}
\end{figure*}

The observable in paleo-detectors is damage tracks caused by nuclear recoils. Neutrinos with energies $E_\nu \lesssim \mathcal{O}(100)\,$MeV give rise to nuclear recoils predominantly via coherent neutral current interactions.\footnote{Additional contributions arise from quasi-elastic charged-current interactions. However, the contributions to the recoil spectrum induced by CC SN neutrinos are suppressed at small neutrino energies by the lack of coherent enhancement, and at large neutrino energies by the quickly falling neutrino flux. The more energetic nuclear recoils which may be induced by high energy neutrinos furthermore lead to longer damage tracks than the nuclear recoils induced by CC SN neutrinos.} The differential recoil spectrum per unit target mass for target nuclei $T$ is given by~\cite{Billard:2013qya,OHare:2016pjy}
\begin{equation} \label{eq:dRdE}
   \left( \frac{\mathrm{d}R}{\mathrm{d}E_R} \right)_T = \frac{1}{m_T} \int_{E_\nu^{\rm min}} \mathrm{d}E_\nu \, \frac{\mathrm{d}\sigma}{\mathrm{d}E_R} \frac{\mathrm{d}\phi}{\mathrm{d}E_\nu} \;,
\end{equation}
where $E_R$ is the nuclear recoil energy, $m_T$ is the mass of $T$, $\mathrm{d}\sigma/\mathrm{d}E_\nu$ is the differential neutral current interaction cross section, and $E_\nu^{\rm min} = \sqrt{m_T E_R/2}$ is the minimum neutrino energy required to induce a nuclear recoil with energy $E_R$. The differential cross section is
\begin{equation}
   \frac{\mathrm{d}\sigma}{\mathrm{d}E_R} (E_R, E_\nu) = \frac{G_F^2}{4\pi} Q_W^2 m_T \left( 1 - \frac{m_T E_R}{2 E_\nu^2} \right) F^2 (E_R) \;,
\end{equation}
with the Fermi coupling constant $G_F$, the nuclear form factor $F(E_R)$, and 
\begin{equation}
   Q_W \equiv \left( A_T - Z_T \right) - \left( 1 - 4 \sin^2 \theta_W \right) Z_T \;,
\end{equation}
where $\theta_W$ is the weak mixing angle and $A_T$ ($Z_T$) the number of nucleons (protons) in $T$. In our numerical calculations, we use the Helm nuclear form factor~\cite{Helm:1956zz,Lewin:1995rx,Duda:2006uk}
\begin{equation}
   F(E_R) = 3 \frac{\sin(q r_n) - q r_n \cos(q r_n)}{\left( q r_n \right)^3} e^{\left( q s\right)^2/2} \;,
\end{equation}
where $q = \sqrt{2 m_T E_R}$ is the momentum transfer and the effective nuclear radius is $r_n^2 \approx c^2 + \frac{7}{3} \pi^2 a^2 - 5 s^2$ with $a \approx 0.52\,$fm, $c \approx \left( 1.23 A_T^{1/3} - 0.6 \right)\,$fm and $s \approx 0.9\,$fm. Note that more refined calculations of the form factors are available, although only for a few isotopes, see e.g. Refs.~\cite{Vietze:2014vsa,Gazda:2016mrp,Korber:2017ery,Hoferichter:2018acd}.

The recoil spectrum, Eq.~\eqref{eq:dRdE}, is converted into a track length spectrum by summing the stopping power $\mathrm{d}E_R/\mathrm{d}x_T$ over all target nuclei $T$ in a material
\begin{equation} \label{eq:dRdx}
   \frac{dR}{dx} = \sum_T \xi_T \frac{dE_R}{dx_T} \left( \frac{dR}{dE_R} \right)_T \;.
\end{equation}
Here, $x$ ($x_T$) is the track length (of $T$), $\xi_T$ is the mass fraction of $T$ in the target material, and the track length for a recoiling nucleus with energy $E_R$ is
\begin{equation}
   x_T(E_R) = \int_0^{E_R} \mathrm{d}E\, \left| \frac{\mathrm{d}E}{\mathrm{d}x_T} \right|^{-1} \;.
\end{equation}
In our numerical calculations, we use the \texttt{SRIM} code~\cite{Ziegler:1985,Ziegler:2010} to calculate the stopping power in composite materials. A more detailed discussion of the calculation of the stopping power (in particular, a description of a semi-analytic calculation of the stopping power and comparison with \texttt{SRIM} results) can be found in Ref.~\cite{Drukier:2018pdy}.

In Fig.~\ref{fig:dRdx} we show the track length spectrum from galactic CC SNe together with background spectra for two minerals, halite (\ccHal) and epsomite [\ccEps].

\section{Backgrounds} \label{sec:Bkg}

We now briefly review the most relevant sources of background for SN neutrino searches with paleo-detectors. These backgrounds are the same as for dark matter searches with paleo-detectors and we therefore refer to reader to Ref.~\cite{Drukier:2018pdy} for a detailed discussion. We note that all the relevant backgrounds stem from nuclear recoils. Natural defects in minerals are either single-site or span across the entire (mono-)crystalline volume and thus do not resemble the damage tracks induced by neutrinos scattering off the nuclei in the target. 

\subsection{Radioactive decays}
\label{sec:radbkg}
The natural minerals used for paleo-detectors will be contaminated by trace amounts of radioactive elements which in turn give rise to background events. Thus, it is crucial to use materials containing as little radioactive contamination as possible. Minerals formed close to the surface of the Earth from the crust's material have prohibitively large uranium-238 and thorium-232 concentrations. Minerals formed in Ultra-Basic Rock (UBR) deposits and Marine Evaporites (MEs) are much cleaner. UBRs and MEs are comprised of material from the Earth's mantle and form at the bottom of evaporating bodies of water respectively. Here, we assume benchmark uranium-238 concentrations of 0.1\,parts\,per\,billion (ppb) in weight for UBRs and 0.01\,ppb for MEs. See Appendix~\ref{app:Uranium} for further discussion.

The most relevant radioactive contaminant in UBR and ME minerals is uranium-238. The half-life of uranium-238 is $4.5\,$Gyr, while the accumulated half-life of all subsequent decays in the uranium-238 decay chain, until it reaches the stable lead-206, is $\sim 0.3\,$Myr. Thus, almost all uranium-238 nuclei which undergo the first decay after the target mineral was formed will complete the decay chain. Due to the kinematics, the most problematic decays are $\alpha$-decays. This is because $\beta$- and $\gamma$-decays lead to the emission of fast electrons, photons, and neutrinos which do not themselves give rise to observable damage tracks in minerals. The associated recoils of the daughter nuclei from these decays are also too soft to produce observable tracks. $\alpha$-decays on the other hand give rise to $10 - 100\,$keV recoils of the daughter nuclei and an $\alpha$-particle with energy of order a few MeV. Here, we assume that the damage track from the $\alpha$-particle itself is not directly observable, though see Ref.~\cite{Drukier:2018pdy} and references therein for a discussion. Thus, the remaining signatures from $\alpha$-decays are the $10-100\,$keV recoils of the daughter nuclei which give rise to damage tracks similar to those induced by scattering of CC SN neutrinos off the target nuclei. However, the typical decays of uranium-238 lead to a complete decay chain, which contains eight $\alpha$-decays. This will lead to eight spatially connected tracks from the various daughter nuclei in the chain. Such signatures are straightforward to distinguish from the isolated damage tracks induced by neutrinos, and we will assume that all such track patterns can be rejected.

However, the second $\alpha$-decay in the uranium-238 decay chain ($^{234}{\rm U} \to {^{230}{\rm Th}} + \alpha$) has a half-life of $0.25\,$Myr. This will lead to a non-negligible population of events which have undergone a single $\alpha$-decay only~\cite{Collar:1995aw}. For minerals with ages long compared to the half-life of uranium-234 and short compared to the half-life of uranium-238, the number of such single-$\alpha$ events per unit target mass is well approximated by
\begin{equation}
   n_{1\alpha} \approx 10^9\,{\rm kg}^{-1} \left( \frac{C_{238}}{0.01\,{\rm ppb}} \right) \;,
\end{equation}
where $C_{238}$ is the uranium-238 concentration per weight in the target sample. The energy of the $^{234}$Th daughter nucleus from $^{238}{\rm U} \to {^{234}{\rm Th}}+ \alpha$ decays is 72\,keV, leading to a population of events with the corresponding (target-dependent) track length indicated by the dash-dotted vertical gray lines in Fig.~\ref{fig:dRdx}. The characteristic track length of such events allows for straightforward modeling of this background, leading to negligible effects on the sensitivity to CC SN neutrinos, as we will see below.

\subsection{Neutron induced backgrounds}
The two dominant sources of (fast) radiogenic neutrons (see the following subsection for cosmogenic neutrons) are spontaneous fission of heavy radioactive elements such as uranium-238\footnote{Note that the tracks from the fission fragments themselves are easily distinguished from neutrino-induced tracks. This is because fission fragments have energies of order 100\,MeV, leading to much longer tracks than the $\lesssim 100\,$keV recoils induced by neutrinos from CC SNe.} and neutrons produced by $(\alpha,n)$-reactions of $\alpha$-particles from radioactive decays with the nuclei in the target sample. Depending on the precise chemical composition of the target sample, either neutrons from spontaneous fission or from $(\alpha,n)$-reactions dominate; we use the \texttt{SOURCES-4A}~\cite{sources4a:1999} code to obtain the neutron spectrum from both sources, including $\alpha$-particles from the entire uranium-238 decay chain. Note that the $(\alpha,n)$ cross sections differ substantially between different elements and isotopes; thus, it is difficult to make general statements. However, light nuclei such as lithium or beryllium display particularly large ($\alpha,n$) cross sections, making minerals containing sizable mass fractions of these elements not well suited for paleo-detectors due to the resulting large neutron fluxes. 

Neutrons lose their energy predominantly via elastic scattering off nuclei, giving rise to nuclear recoils that are indistinguishable from those induced by neutrinos. Because of the mismatch between the neutron mass and those of most nuclei, neutrons lose only a small fraction of their energy in a single scattering event. For example, a $\sim 2\,$MeV neutron would give rise to $\sim 200$ nuclear recoils with $E_R \gtrsim 1\,$keV in a target material comprised of $m_T \sim 100\,$GeV nuclei. This background is highly suppressed in target materials containing hydrogen: since neutrons and hydrogen nuclei (i.e. protons) have approximately the same mass, neutrons lose a large fraction of their energy in a single collision with a hydrogen nucleus. Together with the relatively large elastic scattering cross section between a neutron and hydrogen, this makes hydrogen an efficient moderator of fast neutrons, even if hydrogen makes up only a relatively small mass fraction of the target mineral. For each target mineral, we compute the nuclear recoil spectrum from the neutron spectra using a Monte Carlo simulation with neutron-nucleus cross sections as tabulated in the \texttt{JANIS4.0} database~\cite{Soppera:2014zsj}.\footnote{We use values from \texttt{TENDL-2017}~\cite{Koning:2012zqy,Rochman:2016,Sublet:2015,Fleming:2015} for the neutron-nucleus cross sections.} The corresponding track length spectrum is indicated by the dotted green lines in Fig.~\ref{fig:dRdx}.

\subsection{Cosmic Ray induced backgrounds}
\label{sec:CRbackgrounds}
Cosmic rays can lead to both nuclear recoils and direct damage tracks in materials, potentially producing background events. However, cosmic ray induced backgrounds can be mitigated by using target materials obtained from deep below the surface of the Earth. The dominant cosmogenic background source will then be neutrons arising from cosmic ray muons interacting with nuclei in the vicinity of the target. Following Ref.~\cite{Mei:2005gm} we estimate the neutron flux to be $\phi_n = \mathcal{O}(100)\,{\rm cm}^{-2}\,{\rm Gyr}^{-1}$ for an overburden of $\sim 5\,$km rock. At a depth of $\sim 6\,$km, the flux is instead $\phi_n = \mathcal{O}(10)\,{\rm cm}^{-2}\,{\rm Gyr}^{-1}$ and for an overburden of $\sim 7\,$km we estimate $\phi_n = \mathcal{O}(0.1)\,{\rm cm}^{-2}\,{\rm Gyr}^{-1}$. We envisage that target samples for paleo-detectors will have masses of order $100\,$g, corresponding to geometric cross sections of $\sim 10\,{\rm cm}^2$. Thus, for minerals obtained from depths larger than $\sim 6\,$km, backgrounds due to cosmic ray induced neutrons will be negligible.\footnote{Note that for depths larger than $\sim 6\,$km, in addition to neutrons from cosmogenic muons, neutron production from atmospheric neutrinos interacting with nuclei in the vicinity of the target must be taken into account as well, see e.g. Ref.~\cite{Aharmim:2009zm}.}

We might also worry about cosmic ray-induced backgrounds in MEs, which form near the surface of the Earth and are buried at a typical rate of a few ${\rm km}/{\rm Myr}$ beneath additional layers of sediment (see for example Ref.~\cite{Schreiber:2000}). However, many accessible ME deposits contain minerals which, at some point in their geological history, have undergone re-crystallization. For example, large ME deposits buried at depths of $\gtrsim 10\,$km can extrude into the more dense overlying rock, in a process known as {\it diapirism}. This process may form structures such as salt domes which are accessible to bore-hole drilling from the surface of Earth. Due to the relatively low density of MEs and higher temperatures at the depths of the original salt bed, the plasticity of the minerals increases, see e.g. Refs.~\cite{Trusheim:1960,Seni:1983,Jackson:1986,Koyi:1993,Vendeville:1991,Jackson:1994,Fort:2012}. This in turn may potentially erase previously recorded tracks, including both the neutrino induced signal tracks and those induced by cosmogenic neutrons. Modeling the effects of such geothermal and geochemical processes on the measured track length spectra in paleo-detectors depends on the specific geological history of the deposit containing the target mineral. However, we will demonstrate that an epsomite paleo-detector could be sensitive to neutrinos from galactic CC SNe even if the effective `mineral age' is significantly less than $1\,$Gyr due to re-crystallization during the process of diapirism.

\subsection{Neutrino induced backgrounds}
\label{sec:nubackgrounds}
Neutrinos from sources other than galactic CC SNe induce nuclear recoils via the same scattering processes as neutrinos from galactic CC SNe. We take the neutrino flux $\mathrm{d}\phi_\nu/\mathrm{d}E_\nu$ for solar and atmospheric neutrinos from Ref.~\cite{OHare:2016pjy}. Our calculation of  the Diffuse SN Neutrino Background (DSNB) is described in Sec.~\ref{sec:Sig} and shown in Fig.~\ref{fig:SN_spec}. There are three separate source regimes for neutrino-induced backgrounds. First, neutrinos with energies $E_\nu \lesssim 20\,$MeV are dominantly produced by solar emission. Second, neutrinos with $20\,{\rm MeV} \lesssim E_\nu \lesssim 30\,$MeV are primarily from the DSNB. Finally, atmospheric neutrinos dominate the flux for larger energies, $E_\nu \gtrsim 30\,$MeV. The corresponding nuclear recoil and track length spectrum, shown by the dashed orange lines in Fig.~\ref{fig:dRdx}, is computed in the same way as for neutrinos from galactic CC SNe. Note that although we will investigate the sensitivity of paleo-detectors to potential variations of the galactic CC SN rate over geological timescales, we keep the background neutrino fluxes fixed at the values which are measured today in our background modeling. Instead, we account for potential variations in the background neutrino fluxes by assuming a large systematic uncertainty on the normalization of the neutrino-induced backgrounds.

It is useful to identify which approximate ranges of track lengths are most relevant for the various neutrino induced backgrounds. In the case of epsomite (right panel of Fig.~\ref{fig:dRdx}) , the $\nu$-induced background is dominated by solar neutrinos for track lengths $x \lesssim 100\,$nm. For track lengths $x \gtrsim 200\,$nm, the $\nu$-induced background spectrum is dominated by atmospheric neutrinos. In the transition region between solar and atmospheric neutrinos, the DSNB is the dominant source of $\nu$-induced background. For other target materials we find similar behavior, but due to different masses of the target nuclei as well as differences in the stopping power, the transitions between the different $\nu$-background components occur at different values of the track lengths. In the case of halite for example, cf. the left panel of Fig.~\ref{fig:dRdx}, the cross-overs occur at track lengths roughly a factor of 2 smaller than what we discussed above for epsomite. 

We will discuss the optimal signal regions in track length space for detecting neutrinos from galactic CC SNe in more detail below, but from the discussion above and Fig.~\ref{fig:dRdx} we can already conclude that the signal region is bounded from below by approximately the track length where the solar and the DSNB neutrino backgrounds cross over. This implies that the normalization of the solar neutrino flux is largely irrelevant to the sensitivity of paleo-detectors to galactic CC SN neutrinos.

\subsection{Background uncertainties}\label{subsec:uncertainties}
In estimating the sensitivity of paleo-detectors, the relevant background quantity is not only the expected number of events in the signal region, but the \textit{uncertainty} on this expected number of background events. While the statistical uncertainty is simply given by the square root of the number of background events in the signal region, we need to make assumptions about the systematic uncertainty on the normalization of each background component which we discuss in the remainder of this section. Note that we use the same values for the systematic uncertainty of each background component as in Refs.~\cite{Baum:2018tfw,Drukier:2018pdy,Edwards:2018hcf}.

Radiogenic backgrounds, including neutrons induced by radioactivity, are well understood and straightforward to calibrate in the laboratory. For example, by placing radioactive sources in the vicinity of a test sample or by studying samples with relatively large concentrations of uranium-238, one can obtain samples with enhanced radiogenic backgrounds. Furthermore, the normalization of radiogenic backgrounds is determined by the concentration of heavy radioactive elements in the vicinity of the target sample and the age of the target sample only. For the real experimental data (rather than calibration data), events with long tracks lengths ($\mathcal{O}(1000)$ nm)  therefore act as a `control region', where the signal is negligible, allowing us to fix the background normalization. One can then extrapolate from the control region to the signal region and thus provide a tight constraint on the expected background. Thus, we assume that radiogenic backgrounds can be well understood and use a $1\,\%$ relative systematic uncertainty on the corresponding normalization.

Neutrino induced backgrounds on the other hand are much harder to characterize; their normalization depends on the flux of neutrinos through the target sample. Although the present day neutrino fluxes are relatively well understood, paleo-detectors would measure neutrino-induced backgrounds integrated over geological timescales, $\lesssim 1\,$Gyr. Furthermore, creating target samples in the laboratory with enhanced backgrounds from neutrinos is challenging since this would require strong neutrino sources with spectra matching the neutrino spectra from the Sun, the atmosphere, and CC SNe. We account for potential time variations in and lack of calibration for the relevant neutrino fluxes by assuming a relatively large systematic uncertainty of 100\,\% in the normalization of the neutrino-induced background spectra. The most prominent neutrino background at the relevant track lengths comes from the DSNB. The flux from the DSNB is thought to follow cosmic star formation history~\cite{Beacom:2010kk,Madau:2014bja} As discussed in Sec.~\ref{subsec:Rtime}, this can be directly measured by observations of high redshift galaxies and is not thought to change by more than $\mathcal{O}(0.1)$, as seen in Fig.~\ref{fig:time}. Our assumed 100\% uncertainty is therefore likely an overestimate, although it has little effect on our sensitivity.

The flux of atmospheric neutrinos may also have changed substantially over geological timescales due to the evolution of a variety of physical systems. While the composition and density of the atmosphere have indeed changed (for discussion, see~\cite{Chemke:2017} and references within), the interaction length for cosmic ray protons, which typically initiate the air showers that produce atmospheric neutrinos, is much shorter than the column depth of the atmosphere at the Earth's surface today~\cite{Battistoni:2002ew,Battistoni:2005}. Thus, only a significantly less dense atmosphere could substantially change the atmospheric neutrino flux and, in such a case, the flux would only decrease. Also, the effects of a changing solar magnetic field on the propagation of cosmic rays through the solar system are modulated on timescales much smaller than $100 \,$Myr (see Ref.~\cite{Muscheler:2016} and references within).

We note that paleomagnetic records and modeling of the geodynamo indicate that the strength geomagnetic field may have varied by an ${\cal O}(1)$ factor over geological timescales (for example, see Ref.~\cite{Driscoll:2016}). However, the strength of the geomagnetic field primarily impacts the comic ray flux by causing changes to the rigidity cutoff\footnote{With rigidity defined as $p_{cr} / Z_{cr}$ for the momentum $p_{cr}$ and charge $Z_{cr}$ of the primary cosmic ray nucleus, the rigidity cutoff indicates the smallest rigidity any particular cosmic ray particle could have without being deflected by the geomagnetic field before interacting with the atmosphere. Over $\sim \,$Gyr timescales, the strength of the geomagnetic field is well approximated by the magnitude of the dipole moment, which is directly proportional to the rigidity cutoff (for example, see Ref.~\cite{Lipari:2000wu}).} and would only change the low energy cutoff of the atmospheric neutrino spectrum at energies where the background is dominated by DSNB neutrinos~\cite{Battistoni:2005}. Regarding the sources of cosmic rays in our galaxy, studies of cosmogenic nuclide\footnote{Cosmogenic nuclides are rare isotopes, for example $^{10}$Be or $^{40}$K, produced in spallation events caused by cosmic rays interacting with the Earth's crust, its atmosphere, or in meteorites which eventually reach the Earth~\cite{Lal:1967}.} data in meteorites and terrestrial samples suggest the cosmic ray intensity within the Milky Way has increased by a factor of $\sim 1.5$ over the last $\sim \,$Gyr (for example, see Ref.~\cite{Wieler:2013}).

Thus, while paleo-detectors could probe several factors which impact the atmospheric neutrino flux over geological timescales~\cite{Jordan:2020gxx}, a systematic uncertainty of 100\,\% in the normalization of the flux as a background is sufficient to project the sensitivity of paleo-detectors to galactic CC SN neutrinos.

\section{Track Reconstruction} \label{sec:ReadOut}

Having described the CC SN neutrino signal and possible backgrounds, we now address some practical aspects of damage track formation and measurement.

We will assume that the entire range of a recoiling nucleus will give rise to an observable damage track. Our studies with \texttt{SRIM} indicate that this is a reasonable assumption for the target materials and recoil energies considered here. However, this assumption must be verified in detailed experimental studies for each combination of target material and read-out method, which are beyond the scope of this work.\footnote{To the best of our knowledge, reliable estimates exist only for the particular case of reconstructing tracks in muscovite mica after cleaving and chemical etching~\cite{Collar:1994mj}.} Further, we assume that low-$Z$ nuclei, in particular $\alpha$-particles (He ions) and protons (H ions) do not give rise to observable damage tracks, and neglect the fading of damage tracks from e.g. thermal annealing. We refer the reader to Ref.~\cite{Drukier:2018pdy} and references therein for a detailed discussion of damage tracks from nuclear recoils and possible read-out methods.

From the track length spectra in Fig.~\ref{fig:dRdx} we can see that the signal-to-background ratio for CC SN neutrino induced events is largest for track lengths of $\mathcal{O}(100)\,$nm. An optimal read-out method therefore requires the resolution to which track lengths can be measured to be $\sigma_x \ll \mathcal{O}(100)\,$nm. This in turn would allow for an accurate measurement of the associated recoil energies. Unfortunately, the feasible size of target samples decreases with increasing spatial resolution. We will assume the use of Small Angle X-ray scattering (SAXs) tomography at synchrotron facilities as our benchmark read-out scenario. SAXs allows for the three-dimensional read-out of bulk samples with spatial resolution of $\sigma_x \sim 15\,$nm~\cite{Holler:2014} and minimal sample preparation~\cite{Schaff:2015}. Note that as yet, damage tracks from ions have not been demonstrated to be reconstructible in three-dimensional SAXs tomography; however, damage tracks have been demonstrated to be observable with SAXs (without prior chemical etching) along the direction of the tracks~\cite{Rodriguez:2014}. While we are proposing a challenging application of SAXs, we estimate that it should be feasible to image $\mathcal{O}(100)\,$g of target material at synchrotron facilities, with spatial resolutions of $\sigma_x = 15\,$nm. Throughout the remainder of the paper we will assume that a mass $M = 100\,$g of target material can be read out with a spatial resolution of $\sigma_x = 15\,$nm. We will also consider track length from 10\,nm to 1000\,nm.

We note, as in Ref.~\cite{Edwards:2018hcf}, that this will present a significant data storage and analysis challenge. Naively, scanning at our assumed level of precision will provide $\sim10^7$ terabytes of data for an $\mathcal{O}(100)\,$g sample. This level of data storage is clearly not feasible. Fortunately, the track lengths relevant for this work are $1\mathrm{\,nm}\lesssim x \lesssim 1000\mathrm{\,nm}$ which would allow for a significant reduction in the necessary data storage by \textit{triggering} on interesting features during the scanning process. For example, storing a cube of data with a diagonal length of $1000\mathrm{\,nm}$ is around $\mathcal{O}(10^2)$ megabytes. Assuming that we have $\mathcal{O}(10^5)$ events in our sample we only need to store ten terabytes of data which can be analyzed more precisely in follow-up studies. The triggering methodology must therefore be both efficient and precise in order to match our requirements that full uranium-238 decay chain tracks (as mentioned in Sec.~\ref{sec:radbkg}) are rejected but signal and background tracks are accepted. Assessing the rejection and acceptance factors are beyond the scope of this work and will be addressed in future publications.

\section{Results} \label{sec:Res}

In this section, we present the projected sensitivity of paleo-detectors to neutrinos from galactic CC SNe. A key parameter which determines the sensitivity of paleo-detectors is the mineral age. Throughout this work, we use the term `mineral age' for the age of the oldest nuclear recoil tracks which persist in the mineral. This should loosely correspond to the time since the (last re-)crystallization of the mineral.

We begin by investigating the minimum time-averaged galactic CC SN rate to which paleo-detectors would be sensitive, both as function of the concentration of uranium-238 in the target sample and as a function of mineral age (Sec.~\ref{subsec:Rmin}). We then investigate the ability of paleo-detectors to decipher the history of galactic CC SNe if one were to study a series of target minerals with ages $100\,{\rm Myr} \leq t_{\rm age} \leq 1\,$Gyr and $\Delta t_{\rm age} = 100\,$Myr. In each individual sample one would deduce the CC SN rate (within experimental uncertainties) integrated over $0 \lesssim t \lesssim t_{\rm age}$. Using a series of samples with different $t_{\rm age}$ then allows one to reconstruct the time dependence of the CC SN rate. In particular, we investigate the extent to which paleo-detectors could be used to measure the time-dependence of the galactic CC SN rate (Sec.~\ref{subsec:Rtime}). Finally we study the sensitivity of paleo-detectors to both a single near-by CC SN and an enhancement of the CC SN rate which is localized in space and time, as would be expected from a starburst event in the Milky Way or the local group (Sec.~\ref{sec:Res_burst}).

Throughout, we use a spectral analysis similar to the procedure used in Ref.~\cite{Edwards:2018hcf} for dark matter sensitivity forecasts. The analysis is performed using the \texttt{swordfish} python package~\cite{Edwards:2017mnf,Edwards:2017kqw}\footnote{\href{https://github.com/cweniger/swordfish}{github.com/cweniger/swordfish}}, which allows for the fast calculation of upper limits and projected confidence regions for reconstructed signal parameters. The main difference to the dark matter analysis is that now the signal arises from damage tracks induced by neutrinos from galactic CC SNe. For all analyses, we consider systematic uncertainties on the normalization of each background component only. See Sec.~\ref{subsec:uncertainties} for a discussion of our assumptions and Appendix~\ref{app:Stat} for further details of our statistical methodology.

To cross check, we also perform a sliding-window cut-and-count analysis analogous to the procedure used in Refs.~\cite{Baum:2018tfw,Drukier:2018pdy}. For the sliding-window cut-and-count analysis, the signal is considered to be within reach when the Signal-to-Noise Ratio (SNR) (the ratio of the number of signal events and the quadratic sum of the systematic and statistical errors of all background components in the signal region) satisfies ${\rm SNR} > 3$. For the spectral analysis, the signal is considered to be within reach if 50\% of experiments would return a $3\,\sigma$ preference for the signal+background hypothesis over background-only~\cite{2012PhRvD..85c5006B}. The significance is evaluated from the Poisson likelihood ratio~\cite{Edwards:2017kqw,Edwards:2017mnf,Edwards:2018hcf}. While the cut-and-count analysis is transparent and intuitive, the spectral analysis is more sensitive.

\begin{figure*}
   \includegraphics[width=0.49\linewidth]{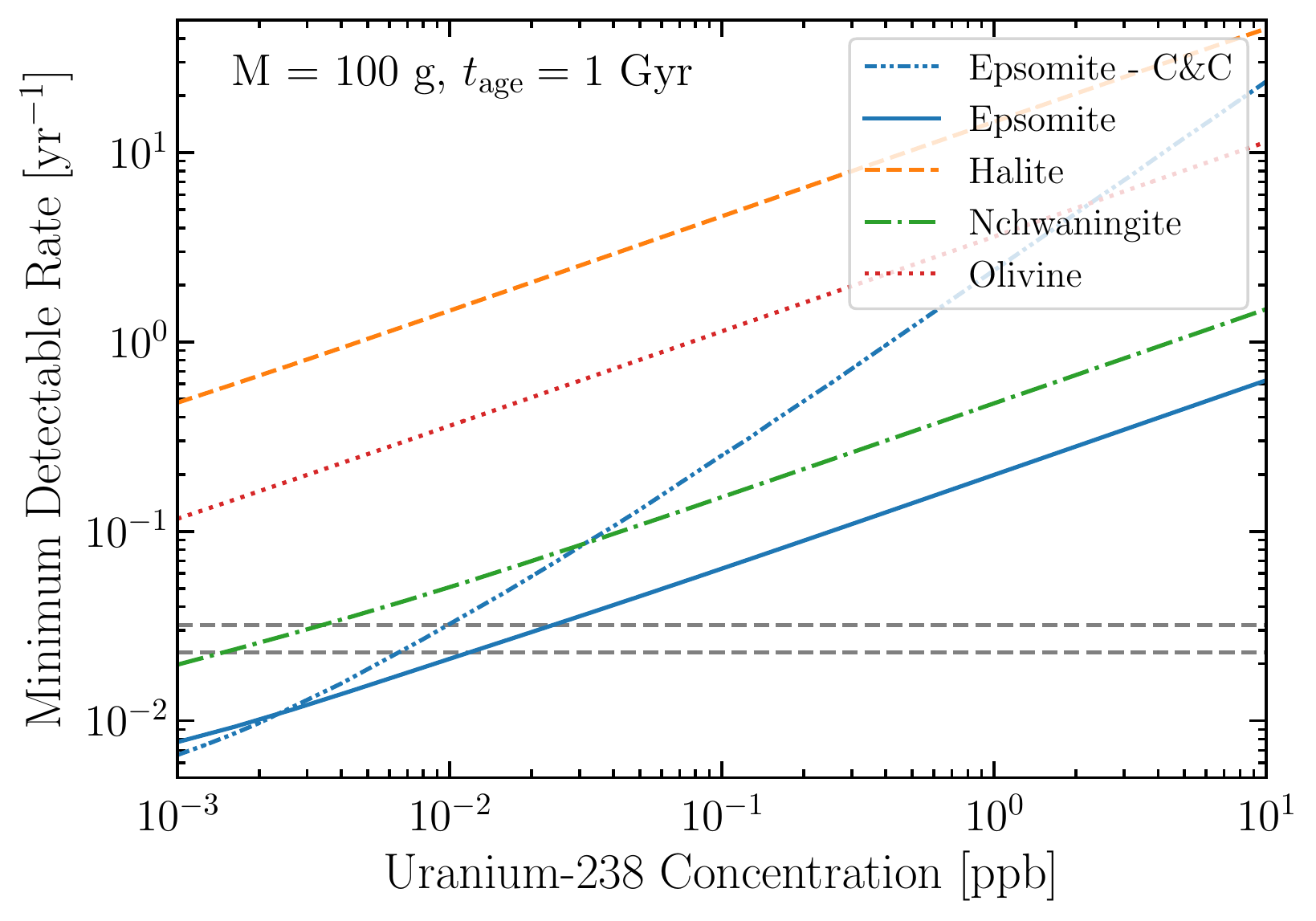}
   \includegraphics[width=0.49\linewidth]{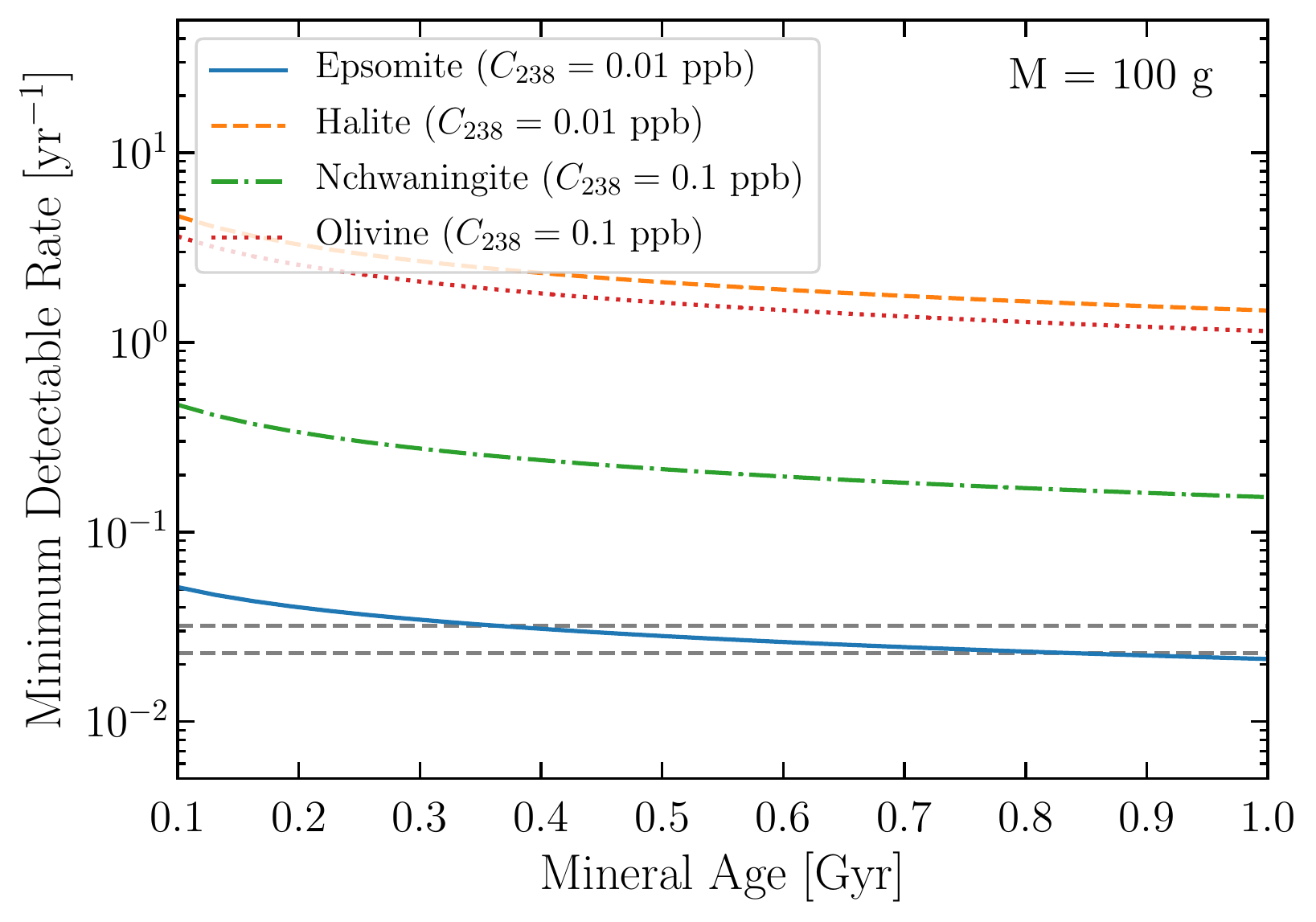}
   \caption{Smallest galactic core collapse supernova rate which could be detected at $3\,\sigma$ in paleo-detectors as a function of the uranium-238 concentration in the target sample (left) and time the mineral has been recording damage tracks (right). Here, we assume that $M = 100\,$g of target material can be read out with a spatial resolution of $\sigma_x = 15\,$nm. The different colored lines are for different target materials as indicated in the legend. For epsomite, we show in addition to the results from the spectral analysis also the sensitivity projections obtained in the sliding-window Cut-and-Count (C\&C) analysis. The horizontal dashed gray lines indicate estimates for the galactic core collapse supernova rate, $\dot{N}_{\rm CC}^{\rm gal} = 2.3 \times 10^{-2}\,{\rm yr}^{-1}$~\cite{Li:2011c} and $\dot{N}_{\rm CC}^{\rm gal} = 3.2 \times 10^{-2}\,{\rm yr}^{-1}$~\cite{Adams:2013ana}. In the left panel we assume that the target sample has been recording damage tracks for $t_{\rm age} = 1\,$Gyr. In the right panel, we assume a uranium-238 concentration of $C_{238} = 0.01\,$ppb in weight for the halite (\ccHal) and epsomite [\ccEps], which are examples of marine evaporites. For olivine [\ccOli] and nchwaningite [\ccNch], minerals found in ultra basic rocks, we assume $C_{238} = 0.1\,$ppb.}
   \label{fig:Rate}
\end{figure*}

From Fig.~\ref{fig:dRdx} it is apparent that the optimal signal region (i.e. where the signal-to-noise ratio is largest) is around track lengths of a few hundred nm. At these track lengths, the signal from galactic SNe is comparable in normalization or somewhat larger than the neutrino induced backgrounds, while the number of background events from radiogenic neutrons is larger than the number of signal events. Here we briefly summarize how the signal can be extracted from the background:
\begin{itemize}
   \item As seen in Fig.~\ref{fig:SN_spec}, the flux from the DSNB is expected to be about two orders of magnitude below the predicted galactic SN flux. As discussed in Sec.~\ref{subsec:uncertainties}, the DSNB flux is not expected to change by more than an $\mathcal{O}(1)$ factor over the lifetime of a paleo-detector. The signal should therefore be clearly identifiable as an excess above the DSNB.
   \item Similarly, from Fig.~\ref{fig:dRdx} and the related discussion in Sec.~\ref{sec:nubackgrounds}, we see that the background from atmospheric neutrinos is typically suppressed relative to the signal by roughly an order of magnitude at the track lengths where the DSNB and atomospheric neutrino induced backgrounds cross over. Although we assume the atmospheric neutrino flux could change by an $\mathcal{O}(1)$ factor over geological timescales, an $\mathcal{O}(100)$ increase in time would be necessary for the atmospheric neutrino background to compare to the radiogenic neutron-induced backgrounds and significantly impact the sensitivity of paleo-detectors to galactic CC SN neutrinos.
   \item The radiogenic neutrons produce a background significantly larger than the signal. Given a 1\% systematic uncertainty on the neutron background, it should be possible to detect a signal which is roughly a factor 100 weaker than the background, as is the case for epsomite in the right panel of Fig.~\ref{fig:dRdx}. Our spectral analysis can go further by tightly constraining the normalization of the background at long track lengths, therefore providing greater sensitivity in the signal region. As we will show, applying either the sliding-window or spectral analysis to an epsomite paleo-detector could allow for sensitivity to galactic CC SN neutrinos.
\end{itemize}

The other backgrounds (and their associated uncertainty) listed above play only a very small role in the spectral analysis and none in the sliding window cut-and-count method. Changing our assumed values of the uncertainty on these remaining backgrounds will therefore not effect our conclusions.

\subsection{Galactic CC SN rate} \label{subsec:Rmin}

In Fig.~\ref{fig:Rate}, we show the minimum galactic CC SN rate which could be observed in paleo-detectors both as a function of the uranium-238 concentration in the target sample (left panel) and the mineral age (right panel). We consider four minerals, halite [\ccHal], epsomite [\ccEps], nchwaningite [\ccNch], and olivine [\ccOli]. Out of these four, epsomite is most promising. This is due to its chemical composition. First, epsomite contains hydrogen, which effectively suppresses the neutron induced backgrounds as described in Sec.~\ref{sec:Bkg}. Further, epsomite does not contain any elements with large $(\alpha,n)$ cross sections and is a ME for which we expect low concentrations of uranium-238, $C_{238} \sim 0.01\,$ppb. Finally, epsomite's particular chemical composition emphasizes the difference between the signal and background spectra: for target nuclei lighter than $\sim 10\,$GeV, i.e.\ lighter than C, the spectrum from SN neutrinos becomes increasingly similar to the background induced by solar neutrinos. For target elements heavier than $\sim 30\,$GeV, i.e.\ heavier than Si, both the signal and background track length spectra become increasingly compressed to shorter lengths. The finite spatial resolution of any given read out method makes it more difficult to distinguish signal from background for such compressed track spectra. In epsomite, the majority of nuclei lie between C and Si in mass, allowing a better separation of signal and background tracks. Thus, we will focus on epsomite as a target mineral for paleo-detectors in the remainder of this paper.

From the left panel of Fig.~\ref{fig:Rate} we find that reading out $M = 100\,$g of epsomite which has been exposed to neutrinos from CC SNe for $t_{\rm age} = 1\,$Gyr should allow for a measurement of the average galactic CC SN rate. Current estimates for the galactic CC SN rate suggest $\dot{N}_{\rm CC}^{\rm gal} \sim 2 \times 10^{-2}\,{\rm yr}^{-1}$ and, as discussed in Sec.~\ref{sec:Bkg}, we expect samples of ME minerals (e.g.\ epsomite) with uranium-238 concentrations of $C_{238} = 0.01\,$ppb to be readily available in nature. 

The right panel of Fig.~\ref{fig:Rate}, where we fix the uranium-238 concentration of epsomite to $C_{238} = 0.01\,$ppb, indicates that measuring the galactic CC SN rate with an $M = 100\,$g epsomite paleo-detector requires target samples which have recorded damage tracks for at least $t_{\rm age} \sim 0.35$ -- $0.8\,$Gyr (depending on the true rate). Note however that if only younger target samples were available, the sensitivity could be recovered by reading out a somewhat larger target sample. This is because the sensitivity depends on the exposure $\varepsilon = M \times t_{\rm age}$; the numbers of signal events and the most relevant background events (i.e. recoils induced by other neutrinos and radiogenic neutrons) scale linearly with $\varepsilon$.

\begin{figure*}
   \includegraphics[height=6.35cm]{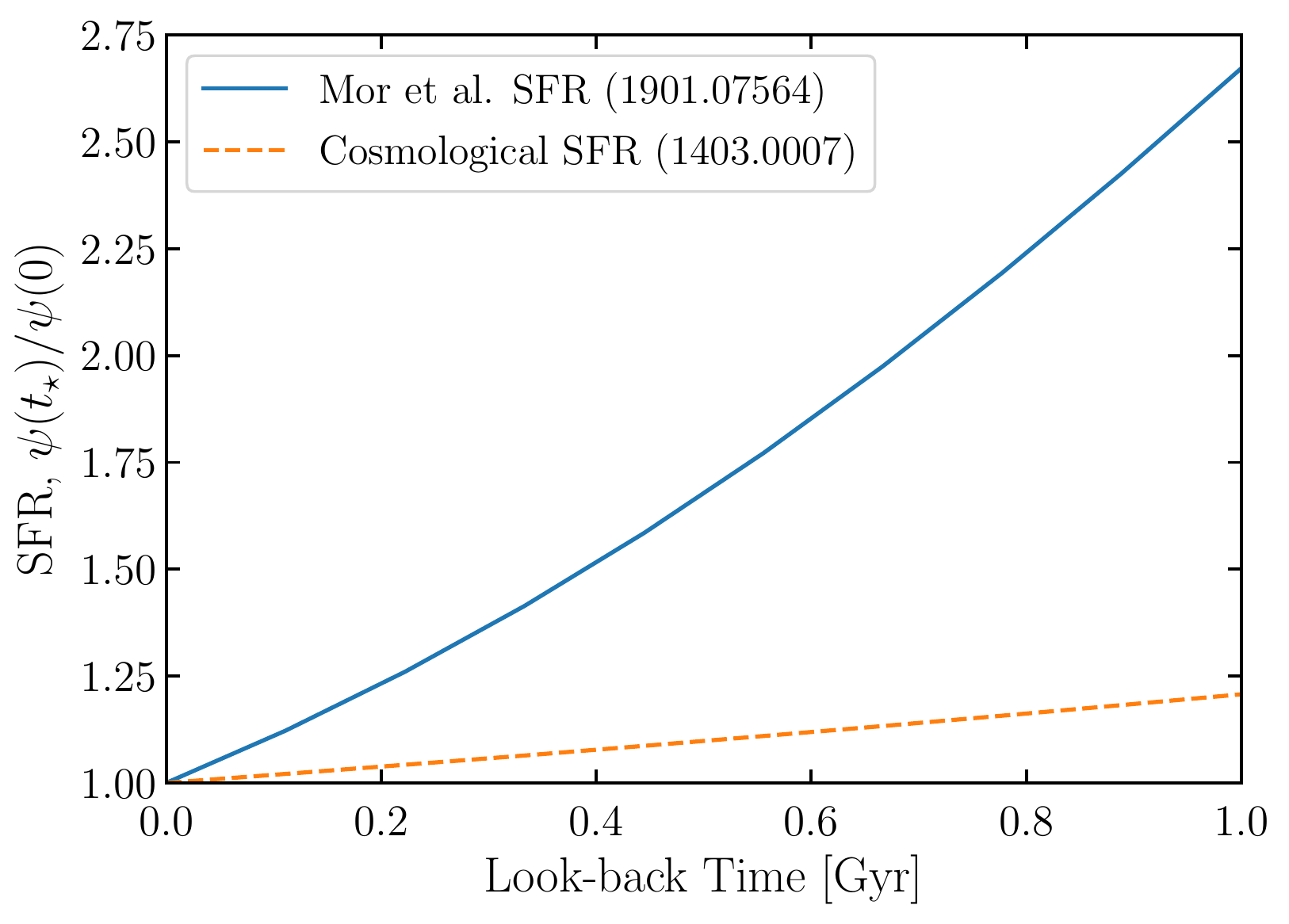}
   \hfill
   \includegraphics[height=6.35cm]{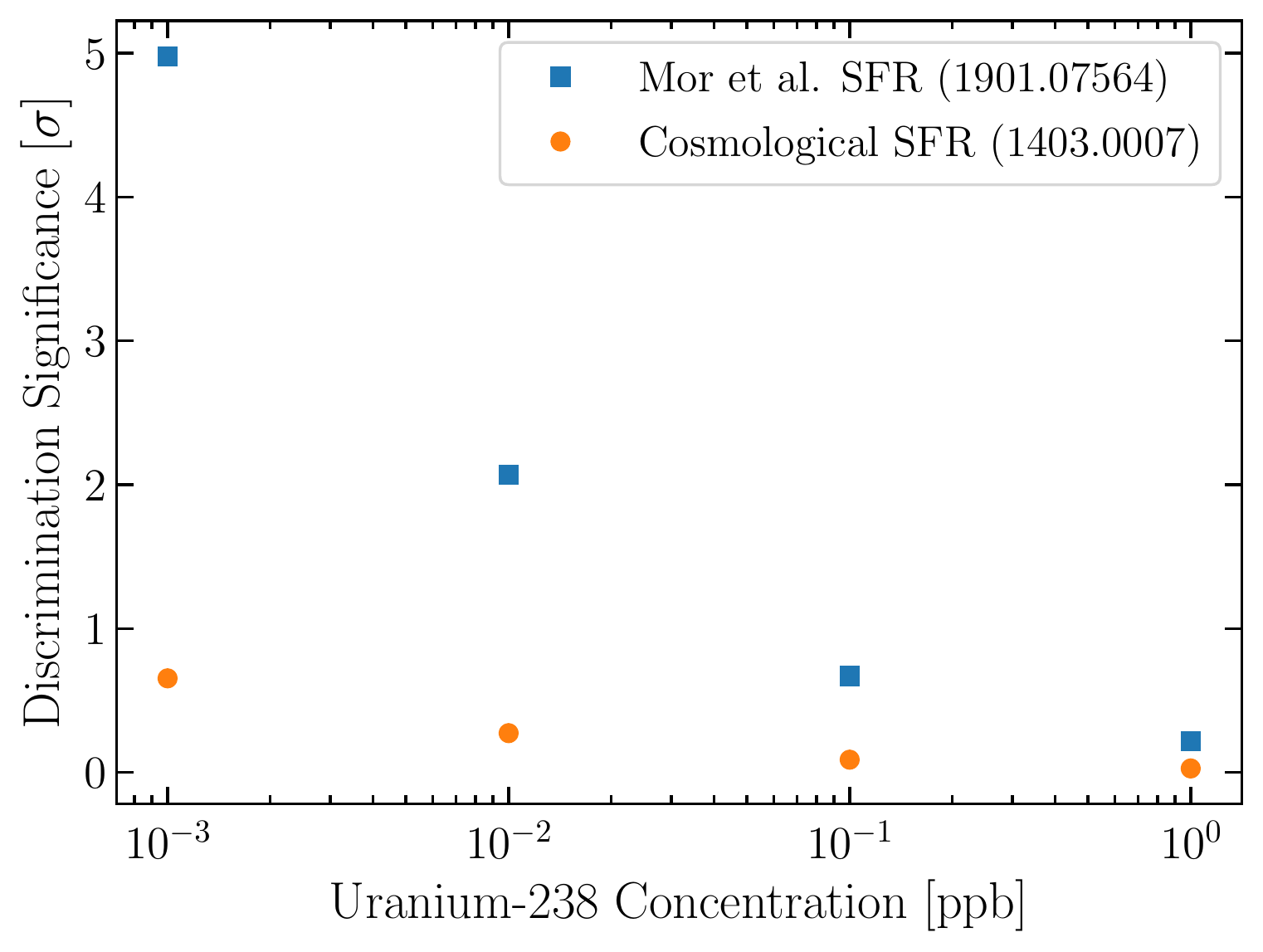}
   \caption{{\it Left:} Benchmark scenarios for the time-dependence of the galactic Star Formation Rate (SFR) $\psi(t_\star)/\psi(t_\star=0)$, as a function of look-back time, $t_\star$, considered in Sec.~\ref{subsec:Rtime}. The CC SN rate $\dot{N}_{\rm CC}^{\rm gal}$ is thought to be directly proportional to the SFR, $\dot{N}_{\rm CC}^{\rm gal} = k_{\rm CC} \psi$, and we use $k_{\rm CC} = 0.0068 M_\odot^{-1}$~\cite{Madau:2014bja}. The blue solid line shows the time-evolution of the galactic SFR of the Milky Way as estimated from Gaia data by~\cite{Mor:2019}, and the dashed orange line the time evolution of the cosmic SFR as estimated by~\cite{Madau:2014bja}. {\it Right:} Assuming different uranium-238 concentrations $C_{238} = \{10^{-3}, 0.01, 0.1, 1\}\,$ppb, we show the discrimination significance with which a time-constant galactic CC SN rate could be rejected if the true galactic CC SN rate evolves with look-back time as the corresponding benchmark scenario shown in the left panel. For both cases, we entertain an experimental scenario where ten epsomite samples with $M = 100\,$g each, which have been recording events for different times $t_{\rm age} = \{0.1, 0.2, 0.3, \ldots, 1.0\}\,$Gyr, are read out with track length resolution of $\sigma_x = 15\,$nm.}
   \label{fig:time}
\end{figure*}

In the left panel of Fig.~\ref{fig:Rate}, we also show, for comparison, the sensitivity forecast from the sliding-window cut-and-count analysis for epsomite.
We note that for uranium-238 concentrations $C_{238} \gtrsim 10^{-2}\,$ppb, the smallest detectable galactic CC SN rate $\dot{N}_{\rm CC}^{\rm gal}$ scales as $\dot{N}_{\rm CC}^{\rm gal} \propto \sqrt{C_{238}}$ for the spectral analysis, while for the sliding-window cut-and-count analysis the scaling is $\dot{N}_{\rm CC}^{\rm gal} \propto C_{238}$. The scaling of the sensitivity in the cut-and-count analysis can be understood from the fact that for $C_{238} \gtrsim \mathcal{O}(0.01)\,$ppb, the sensitivity is limited by the systematic error on the number of background events induced by radiogenic neutrons in the signal region. For the spectral analysis, the error on the number of background events can be reduced by making use of control regions at longer track lengths. The error on the number of background events in the signal region then scales as $\sqrt{C_{238}}$, since it is given by the statistical error on the number of background events in the control regions. For $C_{238} < \mathcal{O}(0.01)\,$ppb, the number of events in the control regions becomes too small to allow such an approach. Simultaneously, the number of background events in the signal region becomes smaller, finally causing both analyses to be limited by the statistical error on the number of background events in the signal region. The remaining differences in the sensitivity are due to the different definitions of sensitivity as discussed above.

Before moving on to estimates of how well the time dependence of the galactic CC SN rate could be constrained by paleo-detectors, it is interesting to ask how precisely the time-averaged CC SN rate could be determined. As before, we consider a benchmark scenario of a $100\,$g epsomite paleo-detector which could be read out with spatial resolution of $\sigma_x = 15\,$nm, e.g.\ by SAXs. We assume that the target mineral has been recording events for $1\,$Gyr and that the true average galactic CC SN rate is $\dot{N}_{\rm CC}^{\rm gal} = 3 \times 10^{-2}\,{\rm yr}^{-1}$. For a uranium-238 concentration of $C_{238} = 0.01\,$ppb, the reconstructed rate could be constrained, at $1\sigma$, to $\dot{N}_{\rm CC}^{\rm gal} = (3.0 \pm 0.7)\times 10^{-2}\,{\rm yr}^{-1}$. For uranium-238 concentrations of $C_{238} = 10^{-3}\,$ppb, the reconstructed rate could instead be constrained to $\dot{N}_{\rm CC}^{\rm gal} = (3.0 \pm 0.3)\times 10^{-2}\,{\rm yr}^{-1}$. Thus, with sufficiently low uranium-238 concentrations it may be possible to constrain the time-averaged galactic CC SN rate to within 10\%, allowing for a discrimination between different estimates in the literature~\cite{Li:2011c,Adams:2013ana}.

\subsection{Time dependence of the CC SN rate}\label{subsec:Rtime}

In the previous subsection we investigated the smallest detectable time-constant CC SN rate. Here and in the following subsection, we instead investigate how paleo-detectors can be used to understand the time-evolution of the galactic CC SN rate.

We entertain two benchmark scenarios for the time-dependence of the galactic CC SN rate: (i) a rate increasing with look-back time according to the best-fit evolution of the galactic Star Formation Rate (SFR) obtained in Ref.~\cite{Mor:2019} from Gaia data, and (ii) a rate increasing with look-back time proportional to the cosmic SFR as parameterized in Ref.~\cite{Madau:2014bja}, cf. the left panel of Fig.~\ref{fig:time}. Note that scenario (i) is based on information from the Milky Way, while scenario (ii) is not (it relies purely on cosmological information). As mentioned in Sec.~\ref{subsec:uncertainties}, the DSNB is thought to roughly track the cosmic SFR.

To quantify the significance at which such scenarios could be distinguished using paleo-detectors, we consider an experimental scenario using ten epsomite samples weighing $M = 100\,$g each, which have been recording events for different times $t_{\rm age} = \{0.1, 0.2, 0.3, \ldots, 1.0\}\,$Gyr. We assume that each target sample is read out with track length resolution of $\sigma_x = 15\,$nm. We then simulate expected signal (and background) events in each sample [for scenarios (i) and (ii)] and calculate the best fit value and error bars for the reconstructed time-integrated CC SN rate in each target sample. Assuming the error bars for the reconstructed rates are described by a Gaussian distribution, we then attempt to fit a time-constant galactic CC SN rate to the mock data, and quantify the statistical significance with which the hypothesis of a constant CC SN rate would be rejected.

In the right panel of Fig.~\ref{fig:time}, we show the statistical significance with which a constant CC SN rate would be rejected in both scenarios as a function of the uranium-238 concentration in the target sample $C_{238} = \{10^{-3}, 0.01, 0.1, 1\}\,$ppb. For scenario (ii), where we assume that the galactic CC SN rate increases with look-back time as the cosmic SFR, we see that it is difficult to distinguish such a time evolution from a constant CC SN rate even if the uranium-238 concentration in the target samples is $C_{238} = 10^{-3}\,$ppb. This is because the cosmic SFR evolves quite slowly in time. For a look-back time of $1\,$Gyr, the cosmic SFR is only increased by a factor of $\psi(t=1\,{\rm Gyr})/\psi(t = 0) \approx 1.2$, using the estimate of the cosmic SFR from Ref.~\cite{Madau:2014bja}. In scenario (i) on the other hand, where the galactic CC SN rate evolves like the estimate for the galactic SFR from Ref.~\cite{Mor:2019}, we find that the hypothesis of a constant galactic CC SN rate could be rejected at more than $3\,\sigma$ if the uranium-238 concentration in the target samples is $C_{238} \lesssim 5 \times 10^{-3}\,$ppb. This is because the estimate for the galactic SFR from Gaia data in Ref.~\cite{Mor:2019} indicates a much faster increase of the SFR with look-back time than the cosmic SFR, $\psi(t=1\,{\rm Gyr})/\psi(t=0) \sim 3$. 

\subsection{Constraining burst-like CC SNe} \label{sec:Res_burst}

\begin{figure*}
   \includegraphics[width=0.49\linewidth]{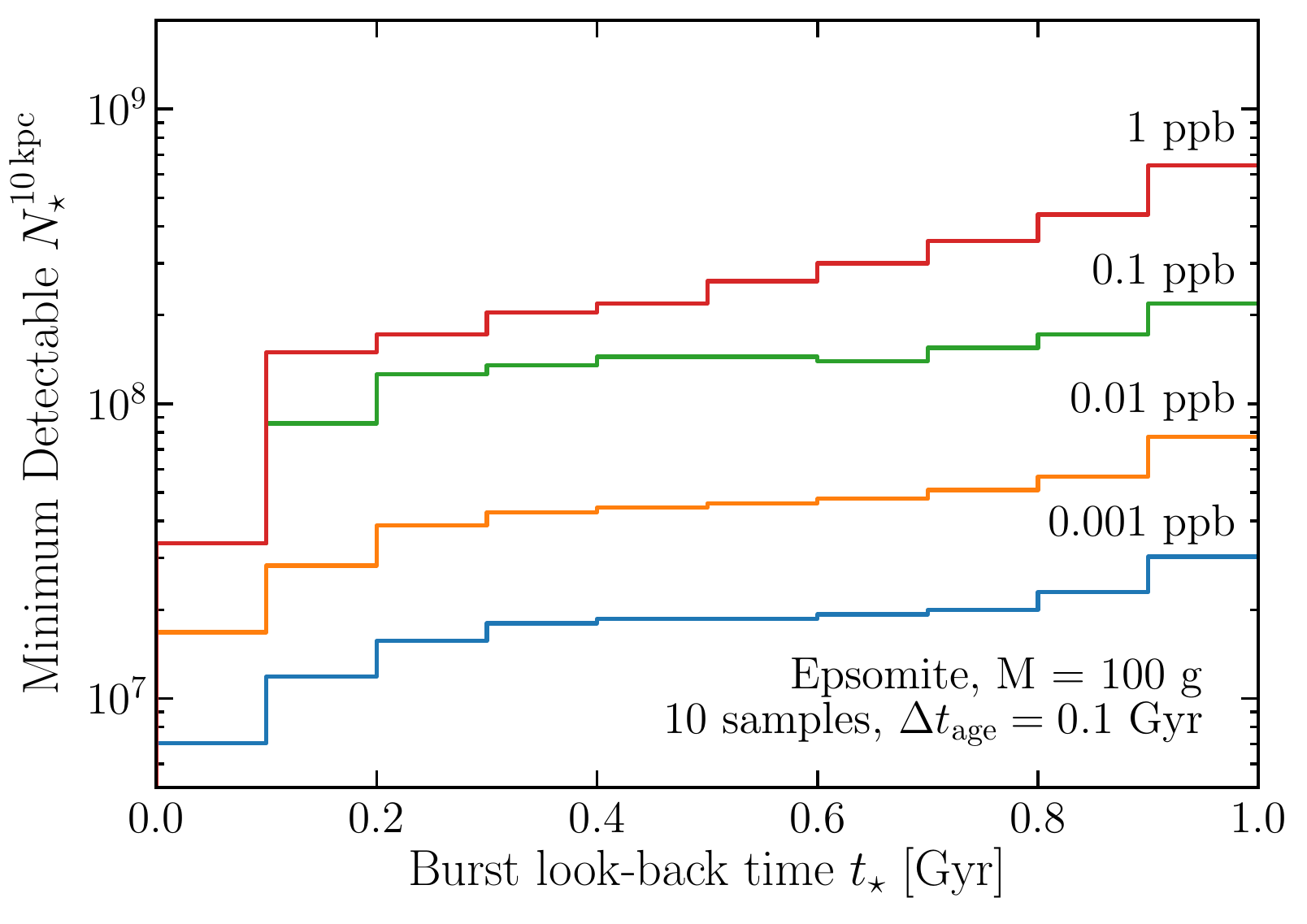}
   \includegraphics[width=0.49\linewidth]{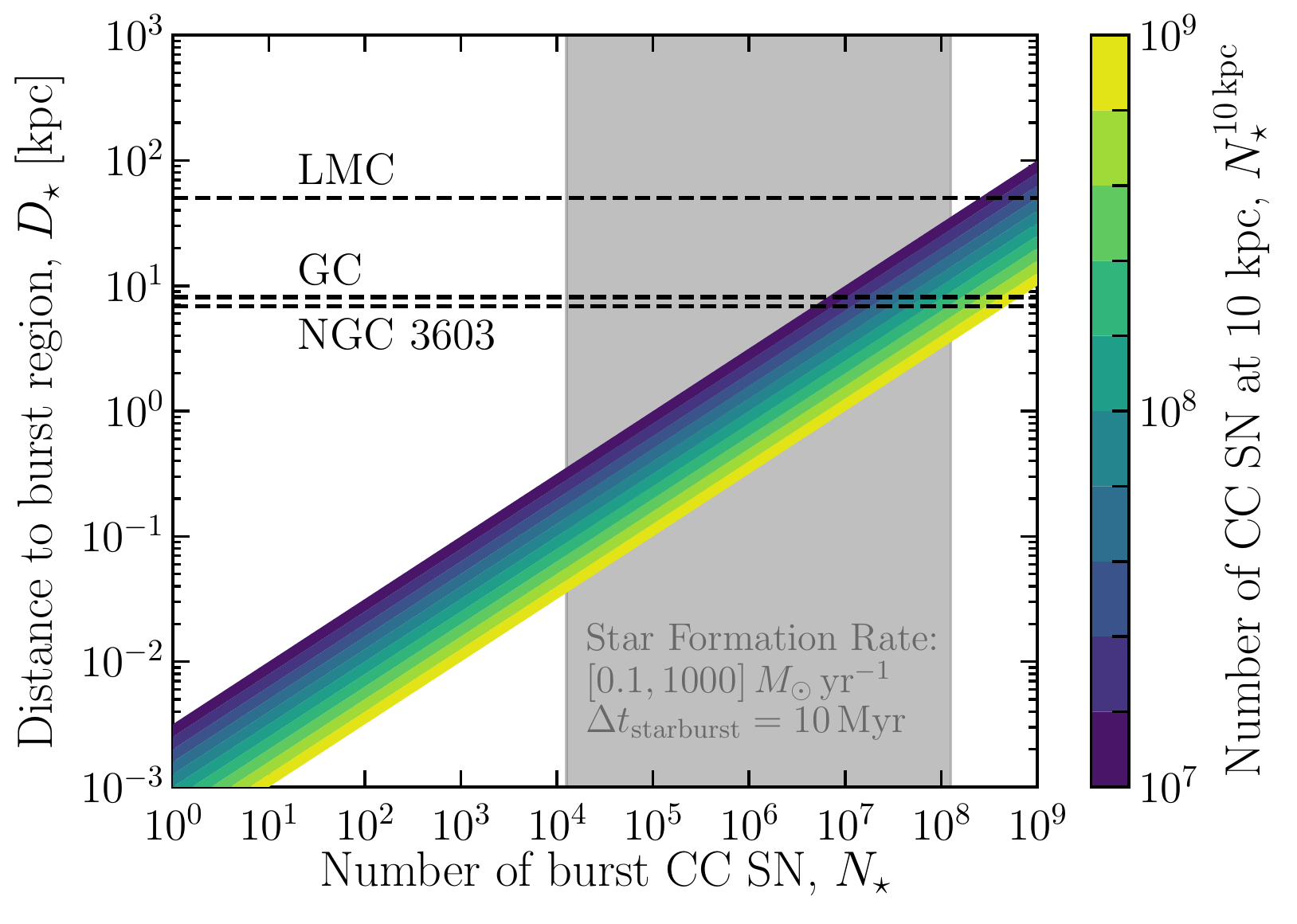}
   \caption{{\it Left:} Smallest number of CC SN events $N_\star$ in a burst-like event at a distance of $D_\star = 10\,$kpc, $N_\star^{10\,{\rm kpc}}$, which could be detected at $3\,\sigma$ with paleo-detectors as a function of the look-back time of the burst-like event $t_\star$. The different colored lines are for different uranium-238 concentrations, $C_{238}$, as indicated by the labels. Throughout, we assume that ten samples of $M = 100\,$g epsomite each, which have been recording recoil events for different times $t_{\rm age} = \{0.1, 0.2, 0.3, \ldots, 1.0\}\,$Gyr, can be read out with track length resolution of $\sigma_x = 15\,$nm. The smallest number of detectable CC SN in a burst like event at any given distance $D_\star$ can be directly obtained by $N_\star \geq N_\star^{\rm 10\,kpc} \left( 10\,{\rm kpc}/D_\star\right)^2$. Similarly, the largest distance at which a single close-by CC SN could be discovered can be obtained by $D_\star \leq 10\,{\rm kpc}/\sqrt{N_\star^{10\,{\rm kpc}}}$. {\it Right:} The colored region indicates the range of $N_\star^{10\,{\rm kpc}}$ within reach of paleo-detectors, cf.\ the left panel, in the plane of the number of CC SNe in the burst-like event, $N_\star$, and the distance of Earth to the burst-like region, $D_\star$. The vertical gray band indicates typical values of $N_\star$ which could occur in a starburst, assuming a duration of the starburst of $10\,$Myr and an average star formation rate of $\psi = 0.1 \div 10^3\,M_\odot$\,yr$^{-1}$. The horizontal dashed lines indicate distances to NGC 2603 (a nebula containing the dense open cluster HD 97950), the Galactic Center (GC), and the Large Magellanic Cloud (LMC), respectively.}
   \label{fig:SB}
\end{figure*}

After investigating the sensitivity of paleo-detectors to a smooth time evolution of the galactic CC SN rate, we now switch to asking if paleo-detectors could be sensitive to time- and space-localized enhancements in the local CC SN rate. The simplest example of such a {\it burst-like} event would be a single near-by CC SN. While such a single near-by CC SN would truly be localized in space and time, an enhancement to the CC SN rate (in a particular region of Milky Way or the local group) for a duration of time significantly smaller than the anticipated timing resolution of $\Delta t = 100\,$Myr would effectively also be a localized event. A starburst event, as described in Refs.~\cite{Kennicutt:1998zb,Genzel:2010na,Negueruela:2014,Ohm:2016abw}, in which the star formation rate (and hence the SN rate) can exceed the average star formation rate of the Milky Way by a factor of $\sim 10^3$ for a period of $\Delta t_{\rm starburst} \lesssim 10\,$Myr, is an example of an effectively localized event which could also be probed by paleo-detectors. 

We parameterize such burst-like events by three parameters, $\{N_\star, D_\star, t_\star\}$. $N_\star$ is the number of CC SNe in the burst-like event, $D_\star$ is the distance to the burst region from Earth, and $t_\star$ is the look-back time to the burst event. For a single close-by CC SN, $N_\star = 1$. For a starburst event, $N_\star$ is given by the average star formation rate $\psi$ during the length of the starburst $\Delta t_{\rm starburst}$ and the number of stars which explode as CC SNe per unit mass, $k_{\rm CC}$. For $k_{\rm CC} = 0.0068\,M_\odot^{-1}$~\cite{Madau:2014bja}, a typical duration of a starburst of $\Delta t_{\rm starburst} = 10\,$Myr, and SFRs of $0.1 \lesssim \psi/(M_\odot\,{\rm yr}^{-1}) \lesssim 10^3$ [see e.g. Ref.~\cite{Kennicutt:1998zb}], we find $N_\star = k_{\rm CC} \psi \Delta t_{\rm starburst} \sim 10^4 \div 10^8$. For reference, the number of CC SNe expected over the entire Milky Way within 1\,Gyr is $2.3 \times 10^{7}$, assuming a constant rate of $\dot{N}_{\rm CC}^{\rm gal} = 2.3 \times 10^{-2}\,{\rm yr}^{-1}$.

To estimate the sensitivity of paleo-detectors to burst events, we follow a similar approach as in the previous subsection. We again assume that ten samples of $M = 100\,$g epsomite detectors have been recording events for different times $t_{\rm age} = \{0.1, 0.2, 0.3, \ldots, 1.0\}\,$Gyr and can be read out with track length resolution of $\sigma_x = 15\,$nm. We simulate mock data, assuming a time- and space-localized injection of additional neutrinos from CC SNe in a burst-like event on top of a constant galactic CC SN rate. Assuming Gaussian errors on the reconstructed CC SN rate in each target sample, we then attempt to fit the null-hypothesis of a time-constant galactic CC SN rate to the mock data and quantify the statistical significance with which this null-hypothesis is disfavored. The number of additional signal events from the burst-like event is proportional to $N_\star/D_\star^2$. We show results for the minimum value of $N_\star/D_\star^2$ required for a $3\,\sigma$ rejection of the null hypothesis as a function of the look-back time to the burst-like event $t_\star$ and the uranium-238 concentration in the target samples $C_{238}$. 

For clarity of discussion we parameterize burst-like events with the three parameters $\{N_\star, D_\star, t_\star\}$, although our analysis is only sensitive to the combination $N_\star/D_\star^2$ as a function of $t_\star$. Hence, our benchmark scenarios of a starburst event or a single close-by CC SN are degenerate, although we discuss the results for both cases separately. We leave the exploration of discriminating such signals for future work. One possibility would be to study the anisotropy of damage track directions. The tracks from an individual close-by CC SN would all arise within $\mathcal{O}(10)\,$s, the duration of the SN neutrino burst. On such time-scales, the target mineral would be virtually stationary in space and hence the signal tracks would have a preferred direction. For a starburst event, the signal events are expected to be generated over a timescale of a few tens of Myr, on which the rotation of the Earth, its orbit around the Sun, and the solar system's movement through the galaxy would wash out the directional preference of the tracks. Note that the directional preference would also allow for additional background suppression when searching for signatures from an individual SN, potentially leading to increased sensitivity.

In the left panel of Fig.~\ref{fig:SB}, we show the minimum number of CC SNe in a burst like event at $D_\star = 10\,$kpc\footnote{Here we use $10\,$kpc as a simple illustration rather than a distance of physical significance.}, $N_\star^{10\,{\rm kpc}}$, for which the null-hypothesis of a time-constant galactic CC SN rate would be disfavored by at least $3\,\sigma$. We show these results for various assumptions on the uranium-238 concentration in the target samples, $C_{238} = \{10^{-3}, 0.01, 0.1, 1\}\,$ppb. Trivially, we find that the smaller the uranium-238 concentration and hence the number of background events induced by radiogenic neutrons, the smaller $N_\star^{10\,{\rm kpc}}$ for which the null-hypothesis of a constant CC SN rate could be rejected. Further, we find that the smaller the look-back time to the burst-like event, $t_\star$, the smaller the number of $N_\star^{10\,{\rm kpc}}$ required to reject the null hypothesis. This is because the same number of signal events induced by the burst-like event would be present in all target samples with $t_{\rm age} < t_\star$, while the number of tracks from the time-constant galactic CC SN rate and from radiogenic neutrons, the most relevant background source, increases linearly with $t_{\rm age}$. 

In the right panel of Fig.~\ref{fig:SB}, we show our results in the $N_\star$--$D_\star$ plane. The colored band indicates different values of $N_\star^{10\,{\rm kpc}}$, the edges of the band approximately correspond to the range of values for $N_\star^{10\,{\rm kpc}}$ that would allow for the rejection of a time-constant galactic CC SN rate for the different assumptions on $t_\star$ and $C_{238}$, as shown in the left panel of Fig.~\ref{fig:SB}. To interpret these results, we indicate a range of values for the number of CC SNe, $N_\star$, in typical starburst events as well as the distance to various regions within the local group where starbursts are likely to occur. We find that for a range of possible starburst parameters, such a burst-like event would be detectable in paleo-detectors if it occurred at a distance corresponding to NGC~6303 (a nebula containing the open cluster HD~97950) or the galactic center. Detection of a starburst-event in the Large Magellanic Cloud (LMC), on the other hand, would require a starburst with a SFR a factor of a few higher than the typical range.

From Fig.~\ref{fig:SB} we can also read off the minimal distance $D_\star$ for which an individual close-by CC SN could be detected with paleo-detectors. Depending on the uranium-238 concentrations in the target samples and the look-back time to the close-by CC SN, the null-hypothesis of a constant galactic CC SN rate could be rejected if the distance to the CC SN was smaller than $D_\star \lesssim 1 \div 10\,$pc. For a spatial distribution of the galactic CC SNe as discussed in Sec.~\ref{sec:Sig} and an average CC SN rate of $\dot{N}_{\rm CC}^{\rm gal}$, the probability that a CC SN has occurred within a distance $D_\star \lesssim 10\,$pc from Earth within 100\,Myr is only $\sim 5\,\%$. However, despite the rather small statistical chance of such an event, close-by SNe are of particular interest since they may be related to mass extinction events, see Refs.~\cite{Ellis:1993kc,Svensmark:2012,Thomas:2016fcp,Melott:2017blc,Melott:2017met,Melott:2019fxa}. Although the time-resolution of paleo-detectors is rather coarse-grained, valuable information about possible close-by CC SNe may still be gained. 

Furthermore, measurements of $^{60}$Fe (and other isotopes produced in CC SNe) in sediments from the Earth and the Moon~\cite{Ellis:1995qb,Fields:1998hd,Benitez:2002jt,Ludwig:2016,Fimiani:2016,Wallner:2016,Schulreich:2018pqg,Fields:2019jtj} as well as the effects of such isotopes on the cosmic ray spectra~\cite{Kachelriess:2015oua,Binns:2016wks,Breitschwerdt:2016,Erlykin:2016jfc} suggest the explosion of at least one CC SN within $D_\star \lesssim 100\,$pc from Earth $t_\star \sim 2 \div 3\,$Myr ago. These claims could be tested with paleo-detectors by studying samples of minerals with $t_{\rm age} \lesssim 10\,$Myr, much younger than what we discussed above. Since in paleo-detectors the signal would arise from the CC SN neutrinos, paleo-detectors would allow for a more direct characterization of these nearby SNe than measurements relying on $^{60}$Fe and similar elements (slowly) propagating in cosmic rays.

\section{Discussion} \label{sec:Discussion}

Paleo-detectors are a proposed experimental technique where one would search for the traces of nuclear recoils recorded in ancient minerals. In minerals that can be used as solid-state track detectors, ions traveling through the crystal lattice give rise to damage tracks which, once created, persist for geological timescales. With modern read-out technology it should be feasible to reconstruct such damage tracks with track length resolutions of order $1\div 10\,$nm. Ions giving rise to such short tracks have kinetic energies $E_R \sim 1\,$keV. Combined with the retention of damage tracks over long times, paleo-detectors would represent a method to probe nuclear recoils down to energy thresholds of order keV whilst obtaining exposures as large as $\varepsilon \sim 100\,{\rm g}\,{\rm Gyr} = 10^5\,{\rm t}\,{\rm yr}$ with current read-out technology. References~\cite{Baum:2018tfw,Drukier:2018pdy,Edwards:2018hcf} explored the potential of paleo-detectors for the direct detection of dark matter. Here, we studied how paleo-detectors can be used to understand galactic Core Collapse (CC) Supernovae (SNe) through the nuclear recoils induced via coherent scattering of neutrinos from CC SNe. 

In Sec.~\ref{sec:Bkg} we discussed the most relevant background sources when searching for recoils induced by neutrinos from galactic CC SNe. At small track lengths (corresponding to less energetic nuclear recoils), the dominant background is solar neutrinos. At larger track lengths (i.e.\ more energetic nuclear recoils) the main background comes from nuclear recoils induced by fast neutrons from the radioactive processes of the trace amounts of uranium-238 and other heavy radioactive elements. Both of these will be present in target materials for paleo-detectors. Note that cosmogenic backgrounds, including neutrons induced by cosmogenic muons interacting in the vicinity of the target materials, can be mitigated by using target samples obtained from depths larger that $\sim 6\,$km rock overburden, e.g.\ from the cores of deep boreholes. Unless the concentration of uranium-238 in the target material is less than $C_{238} \lesssim 10^{-14}$ in weight, the sensitivity of paleo-detectors to neutrinos from galactic CC SNe will be limited by radiogenic neutrons. As discussed in detail in Appendix~\ref{app:Uranium}, we expect to be able to find target samples with uranium-238 concentrations of $C_{238} \sim \mathcal{O}(10^{-11}) = \mathcal{O}(0.01)\,$ppb.

For these concentrations of heavy radioactive elements, we showed in Sec.~\ref{sec:Res} that one could measure the time-averaged galactic CC SN rate using paleo-detectors if the true rate is within the range of current estimates of $2 \div 3$ CC SNe per century in the Milky Way. We also investigated how paleo-detectors could be used to understand the time-evolution of the galactic CC SN rate. To this end, we considered an experimental scenario where one would use ten target samples which have been recording nuclear recoil tracks for different times $t_{\rm age} = \{0.1, 0.2, 0.3, \ldots, 1\}\,$Gyr. If the galactic CC SN rate was a factor of $\sim 3$ higher 1\,Gyr ago than today, as indicated by Gaia data~\cite{Mor:2019}, paleo-detectors would allow one to reject the hypothesis of a time-constant galactic CC SN rate to high statistical significance. If, on the other hand, the galactic CC SN rate increases with look-back time similarly to the cosmic star formation rate (corresponding to a galactic CC SN rate a factor $\sim 1.2$ higher 1\,Gyr ago than today) the data obtainable through this experimental scenario would not suffice to distinguish such a time evolution from a time-constant galactic CC SN rate. 

Finally, we investigated how paleo-detectors could be used to learn about an enhancement of the local CC SN rate on timescales short compared to the time resolution of paleo-detectors, which is of order 100\,Myr. Such a {\it burst-like} enhancement of the CC SN rate could arise from a single close-by CC SN, or from a period of significantly enhanced star formation activity in some region of the local group, i.e. a starburst period. For the latter, we have demonstrated that paleo-detectors could detect a sizable starburst period in the galactic center or a region of our galaxy of comparable distance, e.g.\ the nebula NGC~3603, if it occurred less than $\sim 1\,$Gyr ago. Paleo-detectors could also be sensitive to a starburst in the Large Magellanic Cloud, but only in the case of an exceptionally strong starburst with star formation rates $\psi \gtrsim 10^4\,M_\odot\,{\rm yr}^{-1}$ sustained for $\Delta t_{\rm starburst} \sim 10\,$Myr.\footnote{Note though that the Large Magellanic Cloud is not large enough to sustain such an exceptionally strong starburst, which would require $\gtrsim 10^{11}\,M_\odot$ of baryonic mass.} Similarly, a close-by individual CC SN during the last $\sim 1\,$Gyr could leave a detectable signature in paleo-detectors if it occurred at a distance of $\lesssim 10\,$pc from Earth. In the analysis carried out here, we only considered the number of nuclear recoils induced by neutrinos from CC SNe. In such an analysis, enhancements in the signal rate from a starburst period or a single close-by SN would be indistinguishable; see Sec.~\ref{sec:Res_burst} for a discussion of how this degeneracy could be broken using the directionality of the signal tracks.

In conclusion, this paper demonstrates that paleo-detectors are a promising experimental technique to obtain information about the rate of galactic CC SNe. The long timescales $t_{\rm age} \lesssim \mathcal{O}(1)\,$Gyr over which paleo-detectors could have recorded nuclear recoils induced by neutrinos from CC SNe would furthermore offer the unique ability for a direct determination of the history of the galactic CC SN rate over geological time-scales. Because the star formation rate is thought to be directly proportional to the CC SN rate, this would allow for a measurement of the star formation history of our galaxy, providing important information for the understanding of the Milky Way. 

\begin{acknowledgments}
We are indebted to A.~Goobar for collaboration in the early stages of this project as well as for comments on the manuscript. We also thank D.~Eichler, M.~Hayes, S.~Nussinov, T.~Piran, and M. Winkler for useful discussions.
SB, PS, AKD and KF acknowledge support by the Vetenskapsr\r{a}det (Swedish Research Council) through contract No. 638-2013-8993 and the Oskar Klein Centre for Cosmoparticle Physics.
SB, PS, and KF acknowledge support from DoE grant DE-SC007859 and the LCTP at the University of Michigan. 
TE, BK, and CW are supported by the NWO through the VIDI research program ``Probing the Genesis of Dark Matter" (680-47-5).
KF is grateful for funding via the Jeff and Gail Kodosky Endowed Chair in Physics.
\end{acknowledgments}

\appendix

\section{Uranium-238 Concentrations} \label{app:Uranium}
The concentration of uranium-238, $C_{238}$, in target samples plays an important role in paleo-detectors because radioactive processes are one of the most relevant sources of backgrounds. As discussed in Sec.~\ref{sec:Bkg}, see also Ref.~\cite{Drukier:2018pdy}, nuclear recoils induced by radiogenic neutrons are of particular relevance. Fast neutrons produced by spontaneous fission and $(\alpha,n)$ reactions lose their energy predominantly via elastic scattering off nuclei within the target material. The mean free path of fast neutrons in typical minerals is $\mathcal{O}(1)\,$cm. Furthermore, fast neutrons undergo $\sim 10 \div 10^3$ elastic interactions before losing enough of their energy to no longer give rise to nuclear recoils similar to those induced by neutrinos from CC SNe or dark matter. 

Because of the range of the neutrons, the relevant uranium-238 concentration is not necessarily in the target volume itself, but rather the average uranium-238 concentration in an $\mathcal{O}({\rm m}^3)$ volume around the target sample. Modeling of radiogenic neutron backgrounds in an inhomogeneous environment would require knowledge of the geometry and composition of the rock surrounding the target samples. In our background modeling, we assume an infinitely-sized mineral of constant chemical composition. Note that inhomogeneities may lead to either higher or lower neutron-induced backgrounds in the target material. For example, the neutron-induced background in a relatively uranium-rich target sample not comprising hydrogen could be lower by orders of magnitude compared to the background calculated using the infinite mineral approximation, if such an $\mathcal{O}({\rm cm}^3)$ target sample was located in a surrounding $\mathcal{O}({\rm m}^3)$ volume of material where the uranium concentration is lower and hydrogen is present. 

Further, we would like to note that the theoretical estimation of uranium concentrations in natural minerals is notoriously difficult. This is because the concentration depends not only on the average uranium concentration of the material in which the mineral forms, but also on the details of how uranium in incorporated into particular minerals during the formation process. For example, many minerals are rather robust to the introduction of heavy radioactive elements into the crystal lattice and thus chemically expel uranium (and similar heavy elements) during their growth. However, the effect of such purification cannot be quantified in general, see e.g. Ref.~\cite{Adams:1959} for a discussion. In the remainder of this section, we will motivate our choices of benchmark values for the uranium-238 concentrations in possible target materials for paleo-detectors. Ultimately, experimental efforts are required to determine the range of uranium concentrations in the most relevant target materials; we are currently coordinating such an effort. Note that once obtained, concentrations of radioactive trace elements in target samples of interest can be measured reliably to levels as low as $\sim 10^{-15}$ in weight, e.g.\ using inductively coupled plasma mass spectroscopy~\cite{Povinec:2018,Povinec:2018wgd}. 

A rather comprehensive discussion of the concentration of uranium-238 and other heavy radioactive contaminants in natural minerals can be found in Ref.~\cite{Adams:1959}. Typical concentrations of uranium-238 in minerals formed from material in the Earth's crust are of the order of parts per million (ppm) in weight, which would lead to prohibitively large numbers of radiogenic background events in paleo-detectors. However, much lower uranium concentrations are found in minerals which compose Ultra Basic Rocks (UBRs)\footnote{Olivine [\ccOli] is very common in UBRs. We also show results for nchwaningite [\ccNch] in this work, a less common mineral found in UBRs which contains hydrogen.} and Marine Evaporites (MEs)\footnote{Halite (\ccHal) is one of the most common MEs. We also present results for epsomite [\ccEps] in this work, a less common example of MEs.}. This is because UBRs form from material in the Earth's mantle and MEs from salt deposits at the bottom of evaporated bodies of water. Both the Earth's mantle and its seas have uranium-238 concentrations a few orders of magnitude below the material in the Earth's crust and therefore minerals in UBRs and MEs are much better suited as paleo-detectors. 

Reference~\cite{Adams:1959} quotes values for uranium concentrations in UBRs of $1 \div 30\,$ppb (parts per billion) and uranium concentrations of $\lesssim 100\,$ppb in MEs. However, the aim of Ref.~\cite{Adams:1959} was not to find the most radiopure rocks; in particular for MEs the quoted values represent upper limits of uranium-238 concentrations. While the ranges of uranium-238 concentrations given in the literature are typically representative of the most likely values for a given UBR or ME, Ref.~\cite{Adams:1959} notes that variations of up to an order of magnitude outside of such ranges are common. While experimental efforts are under way to better characterize the distributions of uranium-238 concentrations in representative target materials for paleo-detectors, in particular for the case of MEs with $C_{238} \lesssim \,$ppb, our benchmark values represent roughly an order of magnitude downward variation from the most likely ranges given in the literature. 

\subsection{Ultra Basic Rocks}

More detailed discussions of UBRs with $C_{238} \lesssim \mathcal{O}(1)\,$ppb can be found, for example, in Ref.~\cite{Condie:1957}. They reported uranium-238 concentrations of $\mathcal{O}(0.1)\,$ppb uniformly distributed in (clino)pyroxenes, minerals which, along with olivine, constitute most of UBRs. Note however that these concentrations can vary upward by a factor of $\sim 100$ from rock to rock, with the upper end of the range of $C_{238}$ consistent with the values reported in Ref.~\cite{Adams:1959}. Reference~\cite{Seitz:1974} found a similar range for the uranium-238 concentration in UBRs, with some minerals having concentrations as low as $\mathcal{O}(0.1)\,$ppb. Further, Ref.~\cite{Seitz:1974} suggests that the large variation of uranium-238 concentration in UBRs stems from different amounts of more uraniferous materials introduced after the original rock had formed. Both Ref.~\cite{Condie:1957} and Ref.~\cite{Seitz:1974} also suggest (but do not conclusively prove) that such alterations are more prevalent in oceanic UBRs than in continental UBRs.

\subsection{Marine Evaporites}

For MEs there is considerably less published data available. In particular, many of the available data sets only provide upper limits on the uranium concentrations since the true level of uranium-238 in MEs is often below the sensitivity threshold of a given measurement technique. Some of the smallest uranium-238 concentrations in MEs have been reported in Ref.~\cite{Thomson:1954}. They reported uranium-238 concentrations of $\mathcal{O}(0.1)\,$ppb in halite, however, their samples exhibited characteristics suggesting significant impurities. A more recent review of trace elements in MEs is given in Ref.~\cite{Dean:1978}. They report uranium-238 concentrations ranging form $\mathcal{O}(0.1)\,$ppb to $\mathcal{O}(10)\,$ppb. Such large variations in uranium-238 concentrations from sample to sample are difficult to explain from first principles. 

However, for MEs one can at least estimate the uranium-238 concentration and demonstrate that this is consistent with the observed range. Additionally, one can therefore estimate the lowest uranium-238 concentrations one may expect to find. Let us consider halite as a typical example of a ME and assume it forms in a body of water large enough such that the environment surrounding the water has a negligible effect on the average uranium-238 concentration (e.g.\ sufficiently deep ocean water). The uniformly distributed uranium concentration in a halite deposit formed under such conditions can be estimated as
\begin{equation}
   C_{238}^{\rm ME} \sim C_{238}^{{\rm H}_2{\rm O}} \times S_{{\rm H}_2{\rm O}}^{-1} \times \alpha_{\rm NaCl}\;,
\end{equation}
where $C_{238}^{{\rm H}_2{\rm O}}$ is the uranium-238 concentration and $S_{{\rm H}_2{\rm O}}$ the salinity of the original water, and $\alpha_{\rm NaCl}$ is the ratio of the uranium concentration in the halite deposit to that of the residue left over from the original water. While our simple approximation does not necessarily hold for MEs formed in shallower bodies of water in which the deposition environment can significantly impact the uranium-238 concentration, we note that the ranges of $C_{238}^{\rm ME}$ values measured in such environments are similar. For typical values of seawater today, $C_{238}^{{\rm H}_2{\rm O}} = 3\,$ppb~\cite{Adams:1959} and $S_{{\rm H}_2{\rm O}} = 35\,{\rm g}\,{\rm kg}^{-1}$, and assuming that uranium from the water enriches the mineral phase of the evaporite and the leftover water residue equally, $\alpha_{\rm NaCl} = 1$, we find $C_{238}^{\rm ME} \approx 90\,$ppb, which is roughly consistent with the upper limit given in Ref.~\cite{Adams:1959}. 

However, much lower uranium concentrations can be accommodated in our estimate. The uranium-238 concentration in seawater does vary and values as low as $C_{238}^{{\rm H}_2{\rm O}} = 0.3$\,ppb have been reported~\cite{Adams:1959}. While the typical uranium-238 concentrations of sea water are not expected to have varied much over the relevant geological times scales, the salinity of seawater is generally assumed to have been significantly higher in the past than it is today, see e.g. Ref.~\cite{Sanford:2013}. Assuming a factor of two increase in the salinity for ancient oceans relative to today, we find $C_{238}^{\rm ME} \sim 4\,{\rm ppb} \times \alpha_{\rm NaCl}$. As discussed e.g.\ in Ref.~\cite{Yui:1998}, typical values of $\alpha_{\rm NaCl}$ are considerably smaller than $\alpha_{\rm NaCl} = 1$ because uranium can be maintained as a stable complex anion in the water residue without being precipitated. Reference~\cite{Yui:1998} reports values of $\alpha_{\rm NaCl} = 0.006$ and $\alpha_{\rm NaCl} = 0.011$ for different samples. Taking such effects into account, one may expect the lower range of typical uranium-238 concentrations in MEs to be $C_{238}^{\rm ME} = \mathcal{O}(0.01)\,$ppb, which we assume as the benchmark value for our background modeling.

\section{Statistical Techniques} \label{app:Stat}

Here we discuss additional details of the spectral analysis used for sensitivity projections. All analyses were performed using \texttt{swordfish} (\href{https://github.com/cweniger/swordfish}{github.com/cweniger/swordfish}), an analysis tool developed in Refs.~\cite{Edwards:2017kqw,Edwards:2017mnf}. \texttt{swordfish} automatically uses the spectral differences between the signal and background models to calculate accurate sensitivity projections, regardless of the statistical regime (Gaussian or Poissonian). This is made possible through the \textit{equivalent counts method}, introduced in Sec.~2.4 of Ref.~\cite{Edwards:2017kqw}. In Sec.~\ref{subsec:Rmin} we calculate the minimum rate required to be detectable at $3\,\sigma$ significance. We define this rate to be the discovery threshold, as discussed in Refs.~\cite{Edwards:2017mnf,2012PhRvD..85c5006B}. In particular, this is given by the value of the rate that leads (in 50\% of the cases) to a rejection of the no-signal hypothesis at $3\,\sigma$. The exact definition is given in Eq.~(8) of Ref.~\cite{Edwards:2017mnf}.

In Sec.~\ref{subsec:Rtime} we discuss the ability of paleo-detectors to decipher the time evolution of the galactic CC SN rate. Here we present the procedure used to calculate the model selection statements in more detail. Note that we proceed similarly for the time varying signal and burst search. Importantly, the ten mineral ages we consider can be treated as independent data sets since no two minerals will record tracks induced by the same neutrino. We first simulate the expected rates, $R^o_i$, from a time varying signal in each age bin $i$. This expectation is specific to the model under consideration. We then calculate the expected errors on the reconstructed values, $\sigma^2_i$, and fit a time-constant rate by minimizing the chi-squared difference,

\begin{equation}
    \chi^2 = \sum_i \frac{\left(R^c_i - R^o_i\right)^2}{\sigma^2_i}\;,
\end{equation}
where $R^c_i$ is the time-constant rate which is varied to best fit $R^o_i$. We then calculate the statistical distinctness between the signals given by $R^o_i$ and $R^c_i$. For this we make use of \textit{Euclideanized signals} $x_i(R)$, a technique developed in  Refs.~\cite{Edwards:2017kqw,Edwards:2018hcf}. The Euclideanized signal method is an approximate isometric embedding of a $d$-dimensional model parameter space (with geometry from the Fisher information metric) into $n$-dimensional Euclidean space here given by : $\vect R \mapsto \vect x(\vect R)$ with $\vect x\in \mathcal{M} \subset\mathbb{R}^n$ and $\vect R \in \mathbb{R}^d$. The full definition is given in Eq.~(A18) of Ref.~\cite{Edwards:2017mnf}. This embedding allows one to estimate differences in the log-likelihood ratio by the Euclidean distance (in units of $\sigma$) as,
\begin{equation}
    d = \sqrt{\sum_i |x_i(R^o_i) - x_i(R^c_i)|^2}\;,
\end{equation}
as shown in Fig.~\ref{fig:time}. Here, $d$ quantifies the degree to which a time-constant rate would be disfavored by a data-set consistent with the time-varying rates we considered. A similar procedure is used in the burst search but instead we compute the minimum enhancement to $R^o_i$ required to give $d>3$, as shown in Fig.~\ref{fig:SB}.

\bibliography{theBib}

\end{document}